\documentclass[aps,prb,reprint,superscriptaddress]{revtex4}

\usepackage{bbold}
\usepackage{graphicx}% Include figure files
\usepackage{dcolumn}% Align table columns on decimal point
\usepackage{bm}% bold math
\usepackage{epstopdf}
\usepackage{color}
\usepackage{longtable} 
\usepackage{amsmath}
\usepackage{amssymb}
\usepackage{sistyle}
\usepackage{mathtools}
\usepackage{bm}
\usepackage{changes}
\usepackage{float}

\newcommand{\yb}{Yb$^{3+}$}
\newcommand{\ybi}{$^{171}$Yb$^{3+}$}
\newcommand{\ket}[1]{|#1\rangle}
\newcommand{\bra}[1]{\langle #1|}
\newcommand{\dfc}{$^2$F$_{5/2}$}
\newcommand{\dfs}{$^2$F$_{7/2}$}
\newcommand{\ybground}{$^2$F$_{7/2}$}
\newcommand{\ybexcited}{$^2$F$_{5/2}$}
\newcommand{\nd}{Nd$^{3+}$}
\newcommand{\er}{Er$^{3+}$}

\newcommand{\eu}{Eu$^{3+}$}

\newcommand{\YSO}{Y$_2$SiO$_5$}

\newcommand{\YVO}{YVO$_4$}

\newcommand{\wnm}{cm$^{-1}$}

\newcommand{\yt}{Y$^{3+}$}
\newcommand{\rate}{s$^{-1}$}
\newcommand{\ybiso}[0]{$^{171}$Yb$^{3+}$:Y$_2$SiO$_5$}

\begin{document}

%Title of paper
\title{Coherence Time Extension by Large Scale Optical Spin Polarization  \\ in a Rare-Earth Doped Crystal}

\author{Sacha Welinski}
\thanks{Present affiliation : Department of Electrical Engineering,
Princeton University, Princeton, NJ 08544, USA}
\affiliation{Chimie ParisTech, PSL University, CNRS, Institut de Recherche de Chimie Paris, 75005 Paris, France}

\author{Alexey Tiranov}
\thanks{Present affiliation : The Niels Bohr Institute, University of Copenhagen, DK-2100 Copenhagen \O, Denmark}
\affiliation{D\'epartement de Physique Appliqu\'ee, Universit\'e de Gen\`eve, CH-1211 Gen\`eve, Switzerland}

\author{Moritz Businger}
\affiliation{D\'epartement de Physique Appliqu\'ee, Universit\'e de Gen\`eve, CH-1211 Gen\`eve, Switzerland}

\author{Alban Ferrier}
\affiliation{Chimie ParisTech, PSL University, CNRS, Institut de Recherche de Chimie Paris, 75005 Paris, France} 
\affiliation{Sorbonne Universit\'es, Facult\'e des Sciences et Ing\'enierie, UFR 933, 75005 Paris, France}

\author{Mikael Afzelius}
\affiliation{D\'epartement de Physique Appliqu\'ee, Universit\'e de Gen\`eve, CH-1211 Gen\`eve, Switzerland}

\author{Philippe Goldner}
\thanks{philippe.goldner@chimieparistech.psl.eu}
\affiliation{Chimie ParisTech, PSL University, CNRS, Institut de Recherche de Chimie Paris, 75005 Paris, France} 
\date{\today}

\begin{abstract}
Optically addressable spins are actively investigated in quantum communication, processing and sensing. Optical and spin coherence lifetimes, which determine quantum operation fidelity and storage time, are often limited by spin-spin interactions, which can be decreased  by polarizing spins in their lower energy state using large magnetic fields and/or mK range temperatures. Here, we show that optical pumping of a small fraction of ions with a fixed frequency laser, coupled with spin-spin interactions and spin diffusion, leads to substantial spin polarization in a paramagnetic rare earth doped crystal, \ybi:\YSO. Indeed, up to more than 90 \% spin polarizations have been achieved at 2 K and zero magnetic field.
Using this spin polarization mechanism, we furthermore demonstrate an increase in optical coherence lifetime from 0.3 ms to 0.8 ms, due to a strong decrease in spin-spin interactions. This effect opens the way to new schemes for obtaining long optical and spin coherence lifetimes in various solid-state systems such as ensembles of rare earth ions or color centers in diamond, which is of interest for a broad range of quantum technologies.
\end{abstract}

%\SW{(@philippe : in this paper we don't show a polarization into only one state. Maybe we can put that result in the SM ?)} 

\maketitle

Systems with both spin and optical transitions offer improved functionalities for quantum technologies \cite{re:2018hh}. They allow storage and entanglement of photonic quantum states for quantum communications \cite{Bussieres:2014dc,Hensen:2015dw}, interfacing  processing nodes with optical networks for distributed quantum computing \cite{Kimble:2008if} or efficient detection for quantum sensing  \cite{Boss:2017gh}. Key parameters in these centers are the  coherence lifetimes of optical and spin transitions, which affect storage time,  operation fidelity and sensitivity. In the solid state, major sources of perturbation to quantum states  are due to magnetic fluctuations in the centers environment that couple to transitions through magnetic dipole-dipole interactions. This magnetic noise is often due to flips of electron or nuclear spins carried by the host material atoms \cite{Zhong:2015bw}, impurities or defects \cite{Serrano:2018ea}, or the centers  of interest themselves  \cite{Bottger:2006jo,Knowles:2013uf}. Considerable work has been devoted to engineer these ensembles, known as the spin bath, in order to reduce their detrimental effect on coherence lifetimes. This includes isotope purification to eliminate elements with non-zero spins \cite{Balasubramanian:2009fu,Tyryshkin:2011fi} and application of high magnetic fields \cite{Bottger:2009ik,Rancic:2017bn} and/or very low temperatures \cite{Kukharchyk:2018ip} to freeze spins in their lower energy state.  Another approach is to decouple the optically addressable spins  from the bath by using clock transitions that are insensitive to magnetic field fluctuations in first order \cite{Fraval:2004cu,Wolfowicz:2013ix,Ortu:2018ig} or by filtering out bath fluctuations by dynamical techniques \cite{Viola:1998ky,PhysRevLett.111.020503}.  Although they can be very efficient, these methods may be complex to implement.

\begin{figure*}
\includegraphics[width=\textwidth]{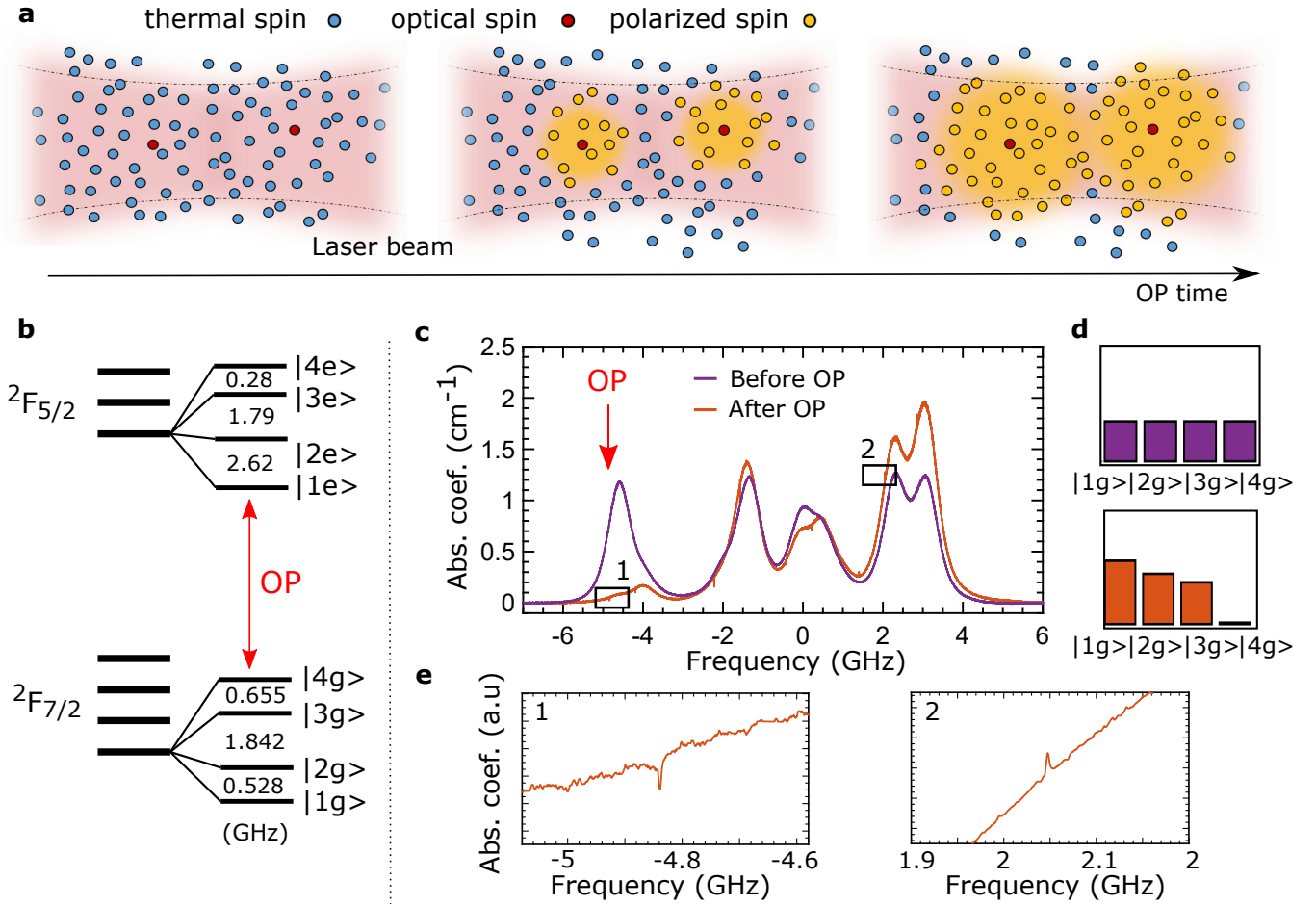}
\caption{Diffusion Enhanced Optical Pumping (DEOP) in \ybi:\YSO. {\bf a} DEOP mechanism: optically pumped spins (red circles) initially surrounded by spins in thermal equilibrium (blue circles) gradually polarize neighboring and further apart spins (yellow circles) through  flip-flop interactions. {\bf b} Energy diagram showing the \ybi\ hyperfine structure. {\bf c} Absorption spectrum without (purple) and after 20 s of optical pumping (orange). The OP is along the  $\ket{4g}\to \ket{1e}$ transition (red arrow in {\bf b,c}). {\bf d} Ground state spin populations $k_{ig}$, normalized by thermal equilibrium values,  over the volume addressed by the laser without OP (top, purple, $k_{ig}=1$) and after OP (bottom, orange, $k_{1g}, k_{2g}, k_{3g}, k_{4g} = 1.67 \pm 0.30, 1.28 \pm 0.36, 1.01 \pm 0.12, 0.04 \pm 0.04$). {\bf e} Enlarged regions 1 and 2 in {\bf c} showing a narrow hole at the laser frequency ($\ket{4g}\to \ket{1e}$) and a corresponding antihole ($\ket{2g}\to \ket{3e}$).}
\label{fig1}
\end{figure*}

Here, we show that optical pumping (OP) of a very small fraction ($\approx 0.5\%$)  of spins in an ensemble  can lead, through spin diffusion, to a  polarization of this ensemble larger than 90 \%. Furthermore, this large scale change in spin populations can be tuned to strongly decrease spin-spin interactions and in turn extend optical coherence lifetimes close to the radiative limit.  This process, which we call Diffusion Enhanced Optical Pumping (DEOP), is illustrated in Fig. \ref{fig1}a. A narrow laser optically pumps a small subset of spins randomly located in the laser excitation volume (labeled 'optical spins' in Fig. \ref{fig1}a, left). The optical spins exchange energy with neighboring spins through flip-flop processes. The latter are initially in thermal equilibrium ('thermal spins'), but because they interact with the strongly polarized optical spins, they become polarized as well ('polarized spins') as shown in Fig. \ref{fig1}a, center. The spin polarization gradually diffuses over the whole optically excited volume through further flip-flop interactions between polarized and thermal spins (Fig. \ref{fig1}a, right). As shown in this Article, DEOP allows coherence lifetime extension by  creating specific, highly out of equilibrium, population distributions among spin levels, a feature difficult to obtain by using magnetic field, low temperature, or by optical pumping alone.

%This large spin polarization is attributed to energy exchanges by flip-flop between $^{171}$\yb{} ground state spins. This process leads to a diffusion of the population imbalance imposed by the optical pumping of a small fraction of the spins. RE ions are randomly distributed over the volume of the crystal and their optical frequencies, determined by local strains, are not expected to be correlated with their location. Ions resonant at a given  frequency are therefore distributed over the volume addressed by the laser, which results, under optical pumping, in a macroscopic spatial spin population gradient and in turn population diffusion, as illustrated  in Fig. \ref{fig1}e. The observation of a decrease of the overall optical absorption, without changes in shape, indicates the absence of correlation between the spin and optical transition frequencies. 

%This process, which we call Diffusion Enhanced Optical Pumping (DEOP),  allows tailoring spin ensemble properties by creating specific, out of equilibrium, population distributions among spin levels, a feature difficult to obtain by  varying  magnetic field or temperature. 
%
%
%over the whole sample volume addressed by the laser, a much larger scale than the one usually obtained with optical pumping alone. This process, which we call Diffusion Enhanced Optical Pumping (DEOP),  allows tailoring spin ensemble properties by creating specific, out of equilibrium, population distributions among spin levels, a feature difficult to obtain by  varying  magnetic field or temperature.

We demonstrate DEOP in a rare earth (RE) doped crystal, \ybi:\YSO\ (YSO). These materials, in which optical and spin coherence lifetime can reach up to ms and hours at low temperatures \cite{Bottger:2009ik,Zhong:2015bw,Goldner:2015ve}, are actively investigated  for spin based quantum photonic applications ranging from quantum memories \cite{Saglamyurek:2011js,Zhou:2012hn,Bussieres:2014dc,Zhong:2017fe,Laplane:2017kw}, processors \cite{Walther:2015hh} and single photon sources \cite{Dibos:2018hh,Zhong:2018ie} to optical-microwave transducers \cite{Williamson:2014fb,Welinski:2019jd}. \ybi:YSO, in which long coherence lifetimes have been shown for both optical and spin transitions at zero magnetic field \cite{Ortu:2018ig}, is particularly promising in this area \cite{Welinski:2016gi,Tiranov:2018kx,Lim:2018hs,Kindem:2018fl}. Using DEOP, we obtained spin polarizations $>90\%$, a much stronger  effect than previously reported in ruby \cite{Jessop:1981jt}. Thanks to its spin ensemble tailoring capabilities, we then used DEOP to reduce or enhance specific spin-spin interactions, and in particular extend \ybi\ optical coherence lifetimes $T_2$ to about 800 $\mu$s, a 2.5-fold increase compared to thermal equilibrium. This is the longest optical $T_2$ reported for any paramagnetic solid state system at zero or very-low magnetic fields, which can be especially interesting for interfacing with superconducting qubits and resonators. DEOP uses a counter intuitive scheme that exploits interactions to ultimately control them and should be effective in  other materials. It paves the way to applications of concentrated, optically active, spin ensembles such as multimode optical or microwave quantum memories and high sensitivity magnetic sensing.

\section{Results}

Experiments were performed using a 10 ppm \ybi:\YSO\ (YSO) single crystal sample (see Methods).  \ybi\ has 1/2 electron and nuclear spins and the corresponding ground (\dfs) and excited state (\dfc) hyperfine structures, for ions in site 2, are presented in Fig. \ref{fig1}b. The optical transition is centered at  978.854 nm (vac.). 
Due to anisotropic Zeeman and hyperfine interactions, all hyperfine levels are non-degenerate and their states show completely symmetric superposition of electron and nuclear spin projections. This results in levels that are insensitive to magnetic field fluctuations at first order under zero external magnetic field. Coherence lifetimes of all transitions are thus significantly enhanced for very low magnetic fields, reaching up to 4 ms  and 180 $\mu$s for spin and optical transitions at 3 K \cite{Ortu:2018ig}.

\subsection{Optical pumping}

Diffusion Enhanced Optical Pumping (DEOP) was studied at 2 K.
A narrow linewidth (about 1 MHz) laser excited the \dfs$\to$\dfc{} transition for a few 10s of seconds. The laser was then blocked for a few ms to let the excited state population relax to the ground state and finally, with a reduced power, shone again on the sample and frequency scanned to determine $^{171}$\yb{} absorption spectrum (see Methods). 
In many crystals, RE spins can be optically pumped at low temperature since excitation to the optical state and subsequent decay often result in population transfer between ground state spin levels.  Since the laser linewidth is usually much narrower than the RE optical inhomogeneous linewidth, optical pumping creates spectral holes and anti-holes in transmission spectra, i.e. regions of low and high absorption that can be as narrow as twice the optical homogeneous linewidth \cite{Macfarlane:1987td,Goldner:2015ve}.  

In \ybi:YSO, spectral hole burning is not the only  phenomenon that occurs. Indeed, as shown in Fig. \ref{fig1}c, after pumping for 20 s the $\ket{4g}\leftrightarrow\ket{1e}$ transition with a 1 MHz linewidth laser, the whole 550 MHz inhomogeneously broadened line vanished. 
This means that essentially all spins in the sample volume addressed by the laser that were initially in the $\ket{4g}$ state ($\approx 2 \times 10^{14}$ spins) have been transferred to other spin states, despite only 0.4 \% of these spins being optically excited. The fraction of excited ions is determined from the overlap between the absorption spectrum and the laser lineshape (see Methods). An analysis of the absorption spectrum, based on previously determined energies and branching ratios of transitions between ground and excited state spin levels, allowed us to accurately determine ground state spin level populations (\cite{Tiranov:2018kx,Businger:2019vw} and see SI for details). This analysis shows that only 4 \% of the  initial  population is left in $\ket{4g}$ (Fig. \ref{fig1}d). We also note in Fig. \ref{fig1}c,e that holes and anti-holes can be observed, although with a low contrast, in the spectrum. They reveal the narrow homogeneous linewidth of the optical transition. The frequency positions of the holes and anti-holes correspond to the pattern expected under spectral hole burning \cite{Tiranov:2018kx}, ruling out spurious effects like large laser drifts during optical pumping.  

As explained in the introduction, we attribute this large spin polarization to energy exchanges by flip-flop between $^{171}$\yb{} ground state spins. This process leads to a diffusion of the population imbalance imposed by the optical pumping of a small fraction of the spins. RE ions are randomly distributed over the volume of the crystal and their optical frequencies, determined by local strains, are not expected to be correlated with their location \cite{Stoneham:1969wq}. Ions resonant at a given optical frequency are therefore distributed over the volume addressed by the laser, which results, under optical pumping, in a macroscopic spatial spin population gradient and in turn population diffusion, as illustrated  in Fig. \ref{fig1}a and SI. The observation of a decrease of the overall optical absorption also indicates the absence of strong correlation between optical and spin transition frequencies.

\begin{figure*}
\includegraphics[width=\textwidth]{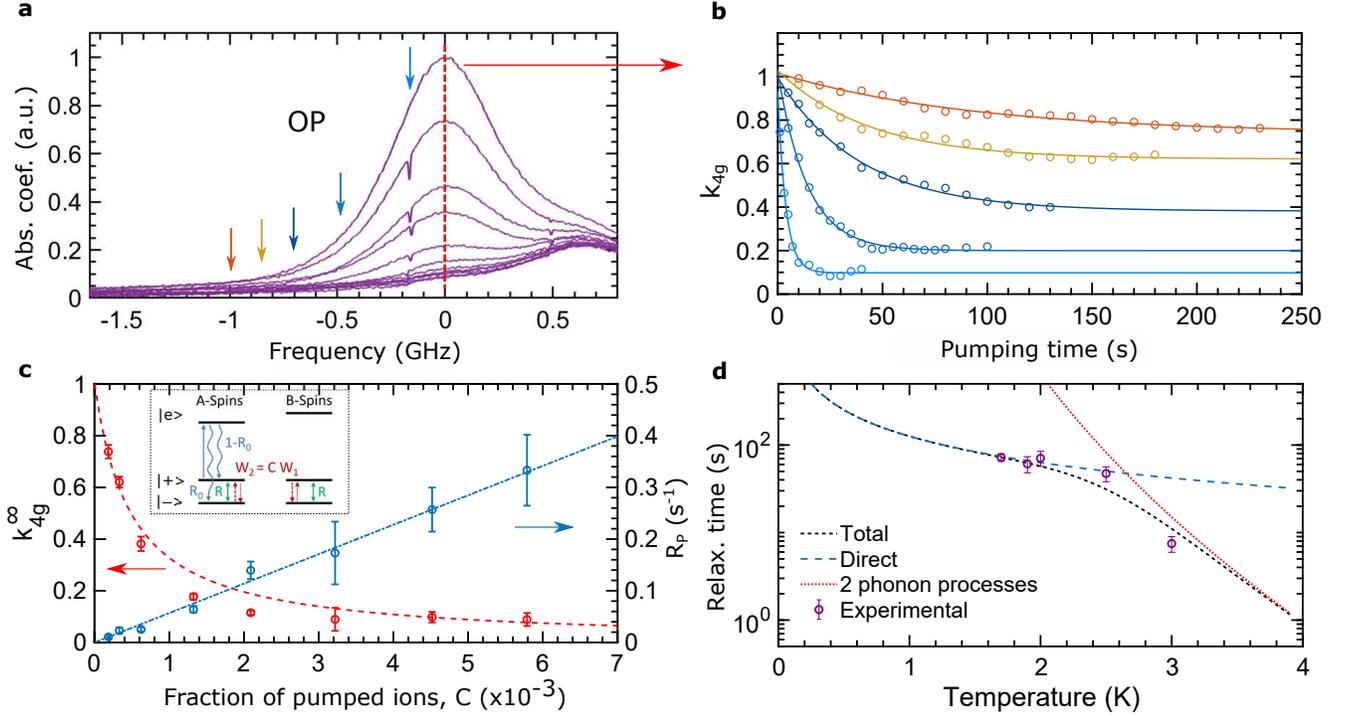}
\caption{Dynamics of DEOP and spin-lattice relaxation. {\bf a} Absorption profile around the $\ket{4g}\leftrightarrow\ket{1e}$ transition after 0 s, 1 s, 3 s, 5 s, 7 s, and every 5 s from 10 s to 40 s of OP at -0.17 GHz (light blue arrow).  The whole inhomogeneously broadened absorption decreases when pumping duration increases. The laser creates a narrow spectral hole at -0.17 GHz clearly seen for pump duration between 1 and 7 s. The small side hole at  +0.5 GHz  originates from the $\ket{3g}$ level and should therefore appears as an antihole. This is explained by a fast relaxation between the $\ket{4g}$ and $\ket{3g}$ levels by flip-flop processes \cite{Tiranov:2018kx}. 
The absorption  centered at +0.66 MHz  corresponds to the $\ket{3g}\leftrightarrow\ket{1e}$ transition. 
{\bf b} Normalized level $\ket{4g}$ population $k_{4g}$ as a function of OP duration for different laser frequencies shown in {\bf a} by color-coded arrows. Solid lines are exponential fits to the data (see text).  {\bf c} Level $\ket{4g}$ normalized steady state populations ($k_{4g}^{\infty}$) and  polarization rate ($R_P$)  as a function of the fraction of pumped ions (see text). Solid lines correspond to fits using a spin 1/2 model and rate equations. Inset: Spin 1/2 model scheme (see text).  {\bf d} Spin lattice relaxation time as a function of temperature deduced from absorption recovery after OP is stopped. Solid lines are fits using direct and two-phonon processes (see text). All error bars correspond to a 95\% confidence interval.}
\label{fig2}
\end{figure*} 

An important goal of this study was to quantify how the degree of polarization depends on the fraction of optically pumped ions, as well as the characteristic time required to reach that polarization. To this end we varied the optical pumping frequency across the $\ket{4g}\leftrightarrow\ket{1e}$ transition  and recorded absorption spectra for different pumping durations $\tau_P$. Fig. \ref{fig2}a shows the region of the absorption spectrum corresponding to the 
$\ket{4g}\leftrightarrow\ket{1e}$ transition, centered at zero frequency detuning, with a smaller contribution from the $\ket{3g}\leftrightarrow\ket{1e}$ transition at +0.65 GHz. The laser frequency is set  at -0.17 GHz, as shown by the hole that appears on spectra recorded for pump durations $1\leq \tau_P \leq$ 25 s.  When $\tau_P$ is increased, the whole inhomogeneously broadened absorption decreases, without change in shape, and reaches a plateau after about $\tau_P=30$ s. From the peak absorption coefficient, measured at 0 GHz in  Fig. \ref{fig2}a, we deduce $k_{4g}$, which is level $\ket{4g}$ population normalized by its value without pumping, i.e. at thermal equilibrium (see Methods and SI). The same experiment was repeated for different laser frequencies, shown by the arrows on Fig. \ref{fig2}a, which effectively reduce the fraction of optically pumped spins. The corresponding variations of $k_{4g}$  are displayed in Fig. \ref{fig2}b.

As the laser moves away from the peak absorption of the  $\ket{4g}\leftrightarrow\ket{1e}$ transition, 
$k_{4g}$ decreases more slowly as a function of $\tau_P$ and plateaus at  a higher  value. We found that it could be well fitted by an exponential expression of the form $k_{4g} = \exp (-R_P \tau_P) + k_{4g}^\infty$. 

 Fig. \ref{fig2}c shows the variation of the steady-state population $k_{4g}^\infty$ and the polarization rate $R_P$ as a function of the fraction $C$ of optically pumped ions, determined  from the overlap between the absorption spectrum and laser lineshape (see Methods).   
The variation of $R_P$ and $k_{4g}^\infty$ can be understood in the following way: when $C$ decreases, each pumped spin   has to polarize a larger number of non-pumped spins to reach a given $k_{4g}$ value; this slows down the overall spin diffusion and decreases the polarization rate $R_P$. 
The degree of achievable spin polarization is limited by the interaction of individual spins with the phonon bath, so-called spin lattice relaxation (SLR), which will counteract DEOP.
Hence, the steady-state population  $k_{4g}^\infty$  is determined by the balance between SLR and spin diffusion rates, such that  a smaller  $R_P$ (and thus $C$) implies a larger $k_{4g}^\infty$.
At the highest fraction of pumped ions, $C = 5.8 \times 10^{-3}$,  $k_{4g}^\infty = 0.08 \pm 0.02$ and  $1/R_P = 3 \pm 0.6$ s, which increases to $k_{4g}^\infty = 0.74\ \pm\ 0.02$ and $1/R_P = 91 \pm 27$ s for  $C = 2 \times 10^{-4}$, the lowest value investigated. Even by pumping such a small fraction of ions, 25\% of $\ket{4g}$ spins are transferred to another level, showing the efficiency of DEOP in this system.

The data in Fig. \ref{fig2}c suggest that $R_P$ depends linearly on $C$ whereas $k_{4g}^\infty$ varies as $1/C$. This can be accounted for by a simple model that treats $^{171}$\yb{} ions as an ensemble of 1/2 spins divided into two groups: the A-spins are  optically pumped to their ground state; the B-spins are not pumped but are expected to polarize to their lower state through DEOP (Fig. \ref{fig2}c, inset). $C$ is therefore the ratio between A and B-spin concentrations. We use rate equations to describe the individual relaxations as well as the flip-flop processes between A and B-spins (see SI for details). They can be solved analytically, leading to 
\begin{eqnarray}
k_{4g}^\infty\equiv p_B^{+,\infty} \approx \frac{1}{ 2+C \beta_{ff}/\beta_o}\label{eq:pb} \\
R_{P} \approx R(2+C\beta_{ff}/\beta_o) \label{eq:rate},
\end{eqnarray}
where $p_B^{+}$ is B-spins  upper state population and $R$ the spin-lattice relaxation rate. These expressions  have indeed the correct dependence on $C$ with respect to experimental observations. $\beta_{ff} = W_1/(W_1+R_o)$ and $\beta_o= R/R_o$, where  $R_o$ is A-spins effective optical pumping rate (Fig. \ref{fig2}c, inset). $W_1$ is the relaxation rate of A-spins by flip-flop with B-spins and $W_2 = C W_1$ is the relaxation rate of B-spins by flip-flop with A spins.

%
%These data were further analyzed by modeling $^{171}$\yb{} ions by an ensemble of 1/2 spins divided into two groups (Fig. \ref{fig2}c, inset): A-spins, in low concentration, are optically pumped; B-spins, the most abundant type, are not pumped. A-spins upper level is assumed to be the one pumped and its population relaxes to the lower level at a rate $R_{o}+R$, where $R$ is the spin-lattice relaxation (SLR) and $R_o$ is an effective optical pumping rate. The lower level population relaxes at rate $R$, which is also the case for B-spins levels.  In addition, A-spin population can decrease by flip-flops with B-spins at a rate $W_1$ and vice-versa at a rate $W_2$. The parameter $C= W_2/W_1$ is equal to the ratio between the numbers of A- and B-spins, i.e. the fraction of pumped ions and  $C \ll 1$ in our experiments. The dynamics of the system is described by rate equations that can be analytically solved (see SI). The steady state population of B-spin upper level population, $p_B^{+,\infty}$ and evolution rate $R_{P}$ under optical pumping of A-spins can be approximated by:
%\begin{eqnarray}
%p_B^{+,\infty} \approx \frac{1}{ 2+C \beta_{ff}/\beta_o}\label{eq:pb} \\
%R_{P} \approx R(2+C\beta_{ff}/\beta_o) \label{eq:rate},
%\end{eqnarray}
%assuming $R_0,W_1 \gg R$, which is well verified in the experiments (see below).  $\beta_{ff} = W_1/(W_1+R_o)$ and $\beta_o= R/R_o$ represent for A-spins branching ratios between flip-flop and total relaxation rates on one hand, and SLR and  optical pumping rate on the other hand. 

As shown in Fig. \ref{fig2}c, reasonable agreement was obtained when fitting experimental $k_{4g}^\infty$ and $R_P$ using Eqs. \eqref{eq:pb} and \eqref{eq:rate}, which indicates that a two-level system can be indeed used to model \ybi\ under these DEOP conditions. Theoretical flip-flop rates show that this is possible due to the fast flip-flops that occur within the $\ket{1g}$ - $\ket{2g}$ and $\ket{3g}$ - $\ket{4g}$ pair of levels (see SI). In this case, each pair of levels can be grouped and considered as one level, leading to an effective 1/2 system.  Flip-flops within others pairs of levels, like $\ket{4g}$ - $\ket{2g}$, are much slower. $W_1$ corresponds to these slow rates, as they are found to be  the limiting interaction for B spins polarization. 
With the additional assumption $R_o\gg W_1$, Fig. \ref{fig2}c fits give  $W_1 = 57\pm 5$ \rate\ and $R$ = (1.4 $\pm$ 0.4) $\times 10^{-2}$ \rate. We estimate $R_o=384 $ \rate\ from excited state lifetime and optical branching ratios, and $W_1 = 13$ \rate\ from narrow hole decays (see SI). These qualitative agreements support our simple spin 1/2 level-rate equation approach. However, we expect that when transitions connecting different ground state levels are simultaneously pumped (see section \ref{coherence}), a more complex 4-level-modeling is necessary.

%"This is possible due to the fast flip-flop relaxation on1g-2g and 3g-4g levels. In this case each pair of levels can be grouped and considered as one level, leading to effective 1/2 system. "
%corresponds to grouping $\ket{1g}$ - $\ket{2g}$ and $\ket{3g}$ - $\ket{4g}$ levels. Indeed fast flip-flops are predicted to occur within these two pairs of levels, whereas
%can successfully account for DEOP in $^{171}$\yb\

We finally recorded the $\ket{4g}\leftrightarrow\ket{1e}$ absorption spectra at different delays after DEOP. As in the previous experiments, the line shape did not change while the initial absorption was gradually recovered, and the peak absorption coefficient allowed us to monitor level $\ket{4g}$ population over all the spins in the volume addressed by the laser. Since flip-flops do not change overall level populations,  the recovery rate $R_c$, obtained by an exponential fit to the data, corresponds to the SLR rate. This is confirmed by $R_c$ temperature dependence shown in Fig. \ref{fig2}d which can be well modeled by a sum of a direct process and 2-phonon processes, with parameters consistent with previous studies at higher magnetic field \cite{Lim:2018hs}, as detailed in SI. At 2 K, the temperature used for DEOP experiments, $R_c = 1/(72$ s) = $1.4 \times 10^{-2}$ \rate, in qualitative agreement with the fitted value $2R =(2.8 \pm 0.8) \times 10^{-2}$ \rate.

\subsection{Optical coherence}
\label{coherence}
\begin{figure*}
\includegraphics[width=\textwidth]{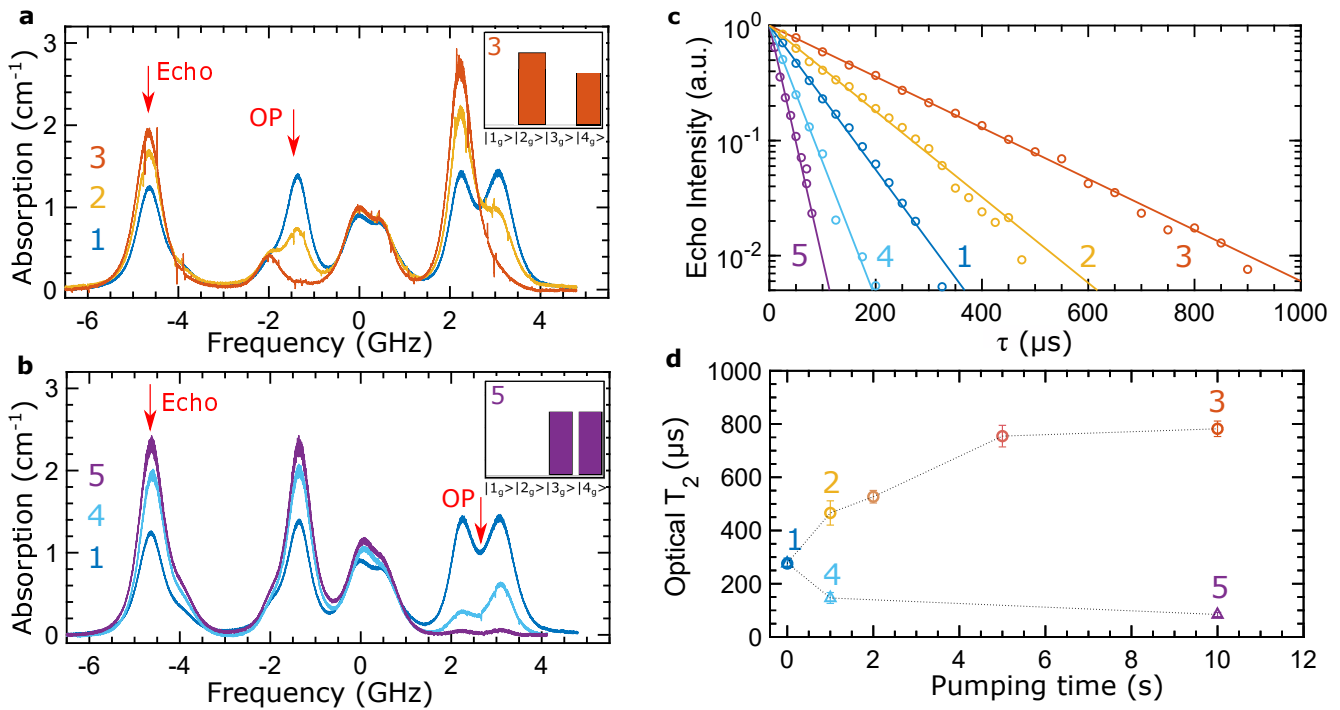}
\caption{Optical coherence lifetime under DEOP. {\bf a} Absorption spectra after 0, 1 and 10 s of OP  at -1.44 GHz. Inset: normalized ground state populations  $k_{1g}, k_{2g}, k_{3g}, k_{4g} = 0.02 \pm 0.02, 2.30 \pm 0.12, 0.02 \pm 0.02, 1.70 \pm 0.06$. {\bf b} Absorption spectra after 0 , 1  and 10 s of OP at  +2.67 GHz. Inset: normalized ground state populations  $k_{1g}, k_{2g}, k_{3g}, k_{4g} = 0.02 \pm 0.02, 0.02 \pm 0.02, 1.95 \pm 0.10, 2.05 \pm 0.08$. {\bf c} Photon echo decays under DEOP for the $\ket{4g}\leftrightarrow \ket{1e}$ transition measured at -4.6 GHz (arrow in {\bf a,b}). Decay colors and numbers correspond to spectra in {\bf a,b}, solid lines are exponential fits to the data. {\bf d} Optical coherence times $T_{2,o}$ deduced from fits in {\bf c} (color-coded).  All error bars correspond to a 95\% confidence interval.}
\label{fig3}
\end{figure*}

We next investigated optical coherence lifetimes, $T_{2,o}$ under DEOP. This was motivated by several studies that have shown that flipping ground state spins of paramagnetic RE can  be a major source of magnetic noise and therefore cause dephasing to RE optical and spin transitions \cite{Bottger:2006jo,Lim:2018hs}. This can be reduced by inducing strong spin polarization under large magnetic field and/or ultra-low temperatures, broadband optical pumping or using excited state spins \cite{Probst:2015ku,Rancic:2017bn,Kukharchyk:2018ip,Cruzeiro:2017hp,Welinski:2019jd}. As shown in the previous section, DEOP also induces large scale spin polarization and could therefore achieve similar effects. 

%However,  DEOP has additional advantages: it is effective even at zero magnetic field, does not need mK range temperatures,  and can be dynamically controlled by adjusting optical pumping frequency, rate or duration.

Optical coherence lifetimes were measured for the $\ket{4g}\leftrightarrow\ket{1e}$ transition under several DEOP conditions. In a first series of measurements, DEOP was performed with a laser set at  -1.44 GHz in the spectrum displayed in Fig. \ref{fig3}a. At this frequency, some ions are pumped along the $\ket{1g}\leftrightarrow\ket{1e}$ transition and others along the $\ket{3g}\leftrightarrow\ket{2e}$ one because of the overlap between these inhomogeneously broadened lines. This results in progressively pumping away the populations of the  $\ket{1g}$ and $\ket{3g}$ levels when the DEOP duration is increased (Fig. \ref{fig3}a). After 10  s, nearly all the  initial populations of these two levels have been transferred to $\ket{4g}$ and $\ket{2g}$.  A second DEOP configuration was studied, setting the laser frequency at +2.67 GHz (Fig. \ref{fig3}c). In this case, the optical excitation is resonant with the four optical transitions $\ket{1g}\leftrightarrow \ket{3e}$, $\ket{1g}\leftrightarrow\ket{4e}$, $\ket{2g}\leftrightarrow\ket{3e}$ and $\ket{2g}\leftrightarrow\ket{4e}$.
As a result, the states $\ket{1g}$ and $\ket{2g}$  are now emptied, and $\ket{3g}$ and $\ket{4g}$ filled, when DEOP is applied (Fig. \ref{fig3}b). These two configurations highlight the versatility of DEOP that allow polarizing spins in different levels and not only in the lowest energy one, as would result from using large magnetic fields or low temperatures. 

For each laser frequency and DEOP duration, the optical coherence time of the $\ket{4g}\leftrightarrow\ket{1e}$ transition was measured with photon echoes (see Methods). The echo decays obtained by varying the delay between the   excitation and rephasing pulses are displayed in Fig. \ref{fig3}c. Decay rates show large variations as a function of DEOP conditions and corresponding coherence lifetimes $T_{2,o}$, obtained by single exponential fits, are gathered in Fig. \ref{fig3}d. Without DEOP, $T_{2,o} = 278 \pm 20$ $\mu$s or  $\Gamma_{h,o}  = 1/\pi T_{2,o} = 1.1  \pm 0.1$ kHz. It reaches 782 $\pm$ 30 $\mu$s after 10 s of DEOP that empties the $\ket{1g}$ and $\ket{3g}$ levels (Fig. \ref{fig3}a,d). The corresponding homogeneous linewidth is $\Gamma_{h,o}  =  407 \pm 15$ Hz. This is the narrowest homogeneous linewidth reported at zero magnetic field for any RE, with the exception of non-Kramers \eu:\YSO{} in which linewidths $< 290$ Hz have been measured \cite{Equall:1994dn}. %\eu{} has however only nuclear degree of freedom and its ground state nuclear hyperfine structure spans only 60 to 160 MHz in this crystal, depending on the isotope \cite{Sun:2005va}. This is about 20 times less than \ybi\ (3 GHz), which in addition has $3 \times10^6$ times stronger spin transition dipole moments \cite{Ortu:2018ig}. \ybi{} is therefore much better suited for interactions with microwave photons while showing comparable optical coherence lifetimes. 

When DEOP is used to empty $\ket{1g}$ and $\ket{2g}$ levels, a very different result is observed: $T_{2,o}$ is strongly reduced, down to 84 $\pm 8$ $\mu$s, equivalent to a homogeneous linewidth  of $\Gamma_{h,o}  = 3.8 \pm 0.4$ kHz. This is a factor of ten difference as compared to the first DEOP configuration, and about 3.5 times the value obtained without pumping. To the best of our knowledge, this is the first demonstration of changes, and especially significant enhancement, in coherence lifetime induced by optical pumping. This is especially significant for systems that should be used at low magnetic field, to take advantage of magnetic insensitive transitions \cite{Fraval:2004cu}, as here in \ybi:\YSO, or when constraints from other devices such as superconducting resonators are relevant.

Contributions to the $\ket{4g}\leftrightarrow\ket{1e}$ homogeneous linewidth can be expressed in terms of levels $\ket{4g}$ and $\ket{1e}$ populations lifetimes $T_{1}$ and pure dephasing $\Gamma_\phi$ as:
\begin{equation}
\Gamma_{h,o} = \frac{1}{2\pi T_{1,4g}}+\frac{1}{2\pi T_{1,1e}}+\Gamma_\phi. \label{eq:coh}
\end{equation}
The excited-state lifetime $T_{1,1e}$ can be taken to be simply its radiative lifetime, $T_{1,1e} = T_{1,o}=$  1.3 ms, hence independent of DEOP, as SLR rates on the same order than in the ground state (close hyperfine and crystal field splittings \cite{Welinski:2016gi,Tiranov:2018kx}) and the spin flop-flop rates are negligible due to the low concentration of excited ions. For the ground state we also disregard SLR contributions to $T_{1,4g}$, as the estimated SLR lifetime is $\approx 2/R_c = 144$ \rate. However, the flip-flop rates can contribute to $\Gamma_{h,o}$ both directly through the $T_{1,4g}$ lifetime and indirectly through the dephasing term $\Gamma_\phi$. The contribution to $\Gamma_\phi$ is then a spectral diffusion process, where flip-flops in the \ybi\ spin bath create a time-varying magnetic field noise on the optically probed ion.

Both the direct and indirect flip-flop contributions are expected to change through DEOP. Our calculations of the flip-flop rates between ions in different hyperfine states show that the highest rates are due to flip flops in between ions in $\ket{1g}$ and $\ket{2g}$, and $\ket{3g}$ and $\ket{4g}$, respectively (see SI). Hence, we expect that the flip flop rate will strongly decrease when ions are pumped into states
$\ket{2g}$ and $\ket{4g}$ using DEOP, as in Fig. \ref{fig3}a. Conversely we expect the flip flop rate to strongly increase when ions are pumped into states $\ket{3g}$ and $\ket{4g}$ using DEOP, as in Fig. \ref{fig3}b. This qualitatively explains the change in coherence time due to DEOP, as seen in Figs \ref{fig3}c and \ref{fig3}d. However, these data are not sufficient to distinguish between the direct (lifetime) and indirect (spectral diffusion) contributions to the coherence time.

To this end, we also performed spin coherence measurements on the $\ket{3g}\leftrightarrow\ket{4g}$ transition at 655 MHz, as described in the SI. For this, we polarized a large fraction of the spins into either the $\ket{1g}$ and $\ket{2g}$ states, or the $\ket{3g}$ and $\ket{4g}$ states, respectively. In both cases the populations in these two levels were essentially the same. An indirect spectral diffusion contribution to the spin coherence lifetime would be roughly equal in both cases, as the flip-flop rates are expected to be the same for both cases (see flip-flop calculations in SI). However, the direct lifetime contribution to the probed states $\ket{3g}$ and $\ket{4g}$ would strongly decrease when spins are polarized into $\ket{1g}$ and $\ket{2g}$ states, as the spin flip-flop probability of spins in both the $\ket{3g}$ and $\ket{4g}$ states would be reduced. 
Indeed, we observed a strong change in spin coherence lifetime for the two cases, going from 0.2 to 2.5 ms as spins are polarized into $\ket{1g}$ and $\ket{2g}$ states.

The spin coherence measurements indicate that direct flip-flop lifetimes significantly contribute to the optical coherence lifetimes in 10 ppm doped \ybi:\YSO. Strong spin polarization using DEOP into selected hyperfine states can strongly reduce this contribution, as well as indirect contributions, to optical and spin dephasing. The data in Fig. \ref{fig3}d show that the DEOP effect is saturated for the longest pumping time, which suggests that direct and indirect contributions  have been largely quenched. This allows to estimate the dephasing contribution to $\Gamma_\phi$ independent from DEOP to 285 Hz. 
 Presumably, \ybi\ in site 1, which represent 50\% of the total \yb{} concentration, cause a significant part of this broadening. It could be reduced to a large extent by using DEOP on these ions using e.g. a second laser. In this case, the remaining dephasing would be interactions with $^{89}$\yt. Since the latter are very slow \cite{Bottger:2006jo}, it could be possible to reach the $T_{2,o}=2 T_{1,o}$ limit.

Although this qualitative analysis can account for the general trends observed, a more detailed modeling and additional experiments are needed to precisely evaluate the processes affecting $T_{2,o}$. In particular, all ground state flip-flops should be included in simulations and their rates determined using spectral hole burning or other techniques. Further theoretical calculations of flip-flop rates and frequency shifts caused by spin flips could also be very useful. 

\section{Discussion}

The spin 1/2 rate equation model  (Eqs. \ref{eq:pb}-\ref{eq:rate}) is convenient to estimate parameters for efficient DEOP, i.e. low remaining population in B-spin upper level, $p_B^{+,\infty}$. First, low values of $\beta_o = R/R_o$ are required and therefore small SLR rate $R$ and/or strong optical pumping, i.e. larger $R_o$. The latter may be limited, as in our case, by the spontaneous emission rate and branching ratios, which in turn can be increased using optical nano-cavities \cite{Zhong:2018ie,Dibos:2018hh,Casabone:2018kc}. Small $R$ values can be achieved by lowering magnetic field and  temperature \cite{Bottger:2006jo}. This is the case in our experiments, running at 2 K and zero magnetic field, giving $\beta_o \approx 3.6 \times 10^{-5}$ and polarizations over 90\%. However, SLR increases quickly with temperature or magnetic field, and Fig. \ref{fig2}d modeling predicts that at 4 K and zero field DEOP polarizes only 11\% of the spins. 
Large $\beta_{ff}$ is also favorable and corresponds to strong flip-flops, i.e. large  $W_1$. This can be obtained with high spin concentration $n_0$ and lower inhomogeneous linewidth  since $W_1\propto n_0^2/\Gamma_{inh,spin}$ \cite{Bottger:2006jo}. In \ybi:\YSO, it is worth noting that the spin linewidth is especially narrow at zero magnetic field, $\Gamma_{inh,spin}= 1$ MHz \cite{Ortu:2018ig}, which increases flip-flop rates and gives $\beta_{ff} \approx 0.13$. Finally, pumping a larger fraction of ions will obviously result in stronger polarization by increasing $C$. However, as demonstrated in our experiments, low $C$ values can still provide strong polarization. 
The case when more than one transition is optically pumped, as investigated in the 'Optical Coherence' section is more difficult to analyze with a simple 1/2 spin model.  Non-pumped spins will interact with several classes of optically pumped spins that are polarized in different levels. This can lead to high polarization, as we observed, but also to opposite population changes and therefore remaining populations in some pumped levels. In this respect, isolated optical transitions, like the $\ket{4g}\leftrightarrow\ket{1e}$ in \ybi:\YSO\ simplify pumping schemes and will appear for spin splittings at least comparable to the optical inhomogeneous linewidth. While this can often be obtained with a high enough magnetic field, it may also increase SLR through direct processes and lower flip-flop rates by increasing spin inhomogeneous linewidth \cite{Veissier:2016cu,Cruzeiro:2017hp}, effects that both reduce DEOP. 

This study suggests that efficient DEOP could be observed  in other paramagnetic RE or transition metal ions doped materials. 
DEOP was observed in Cr$^{3+}$:Al$_2$O$_3$ \cite{Jessop:1981jt} and can also be seen in \nd:\YVO{} as shown in Fig. 5 of Ref. \cite{Afzelius:2010id}, although it was not recognized as such in this work. It could especially be observed in other candidates of interest for applications in quantum technologies including \er, although this ion suffers from inefficient optical pumping which increases the requirement on low SLR \cite{HastingsSimon:2008fj}.  Paramagnetic RE with non zero nuclear spins that show ground state splittings of a few GHz at zero magnetic field such as $^{167}$\er, $^{145}$\nd\ or $^{173}$\yb\ could also behave similarly to \ybi\ for DEOP. Finally, other concentrated spin systems with optical transitions, such as NV$^-$ centers in diamond, could also show DEOP. 

%, with sufficiently narrow spin inhomogeneous linewidths, large enough dopant concentration,
%as well as large Zeeman or hyperfine splittings at low magnetic fields. 

Using DEOP, we managed to reduce by 250\% the optical homogeneous linewidths, to get a value of $\Gamma_{h,o}  =  407 \pm 15$ Hz, which is the narrowest homogeneous linewidth reported at zero magnetic field for any RE, except for \eu:\YSO{} \cite{Equall:1994dn}.
However,  \eu{} only possesses  nuclear degrees of freedom and its ground state nuclear hyperfine structure spans only 60 to 160 MHz in this crystal, depending on the isotope \cite{Sun:2005va}. This is about 20 times less than \ybi\ (3 GHz), which in addition has $3 \times10^6$ times stronger spin transition dipole moments \cite{Ortu:2018ig}. \ybi{} is therefore much better suited for interactions with microwave photons while showing comparable optical coherence lifetimes.

As it is the case for \ybi:\YSO, DEOP could extend optical and/or spin coherence lifetimes of other solid state systems by reducing spin-spin interactions. It therefore allows keeping a high concentration of active species with low dephasing. This is a particularly important point for ensemble based quantum devices, like absorptive quantum memories that require high optical absorption \cite{Bussieres:2013br}. In addition, various configurations of populated levels can be in principle obtained, allowing to select the best configurations for e.g. strong optical and spin transitions, long coherence lifetimes, long lived shelving states etc. DEOP can also provide large scale spin initialization prior to processing and/or spectral tailoring. As an example, we achieved $96 \pm 1 \%$ polarization in the single $\ket{4g}$ level of \ybi:\YSO\ (see SI). Finally, in the case of systems with different sites for optically adressable spins, a common feature in rare-earth doped crystals, it can also lower the perturbations from the unused centers.

In conclusion, we have observed large-scale spin polarization under laser excitation at fixed frequency in a rare earth doped crystal, \ybi:\YSO. This is explained by a combination of optical pumping and spin diffusion by flip-flops that results in $>90$\% polarization for all spins in the sample volume addressed by the laser. The efficiency and versatility of the process is furthermore demonstrated by significantly increasing and decreasing optical coherence lifetimes $T_{2,o}$, depending on the pumping conditions. The longest $T_{2,o}$ recorded, $\approx 800 \mu$s, is the longest recorded for a paramagnetic RE at zero magnetic field and is comparable to values for non-paramagnetic RE. Given the other favorable optical and spin properties of \ybi:\YSO, our results open the way to new designs for broadband and efficient quantum memories for light \cite{Businger:2019vw} or optical to microwave transducers. We expect this process to be efficient in other rare earth doped crystals and concentrated systems of optically addressable spins like color centers in diamond. It could be used to tailor spin baths and therefore extend coherence lifetimes, or initialize  spins on a large scale, topics which are central to many quantum technologies.

\section{Acknowledgements}

This project has received funding from the European Union's Horizon 2020 research and innovation programme under grant agreement No 71272, the IMTO Cancer AVIESAN (Cancer Plan, C16027HS, MALT) and FNS Research Project No 172590.

\section{Author contributions}

S.W. and A.F. grew the sample, S.W. and P.G. conceived and performed the experiments, except for the additional spin coherence measurements which were performed by M. B. and A. T.. S.W., P.G., M.A., A.T. and M.B analyzed the results, and S.W. and P.G. wrote the manuscript with inputs from all co-authors. P.G. and M.A. provided overall oversight of the project.

\section{Methods}

\subsection{Sample}
The \YSO\ single crystal was doped with 10 ppm of \ybi\ (94\% isotopic purity, see SI) and grown by the Czochralski technique. It was cut along the extinction axis $b$, $D_1$ and $D_2$, with  light propagating along the $b$ axis and polarized along $D_2$ for maximal absorption. The length of the sample along $b$ was 9.4 mm. \YSO\ has a monoclinic structure and belongs to the C$^6_{2h}$ space group. \yb\ can substitute \yt\ ions equally in their two sites of $C_1$ point symmetry. 

\subsection{Experimental setup and optical pumping}
The sample was placed inside a liquid helium bath cryostat at 2 K. Excitation was provided by a tunable single mode diode laser (Toptica DL 100) with a spectral width of 1 MHz. The beam on the sample was weakly focused on the sample with a diameter  of 1 mm. All experiments were performed in transmission mode. Spectra were calibrated by recording signal from a Toptica FPI 100 Fabry-Perot interferometer (1 GHz free spectral range).  An acousto-optic modulator (AOM, AA Optoelectronics MT80) in single pass configuration was used to gate the laser. The detector was an amplified Si photodiode (Thorlabs PDA150A). The power during optical pumping and scans was 7 mW and 0.4 mW, with frequency scans performed at a rate of 3 GHz/ms. A delay of 10 ms was kept between optical pumping and scanning to let the excited state population decay to the ground state. To probe absorption recovery, scans were performed at different delays after 20 s of optical pumping and at different temperatures.  

Fraction of pumped ions are calculated from absorption spectrum using the formula 
\begin{equation}
C = \frac{2}{\pi}\frac{\Gamma}{\Gamma_0}\frac{\alpha}{\alpha_0},
\end{equation}
where $\alpha$ and $\alpha_0$ are absorption coefficients at the laser frequency and peak of the line, $\Gamma_0$ the full width at half maximum and $\Gamma$ the pumped region spectral width. We have $\Gamma_0 = 550$ MHz \cite{Ortu:2018ig} and $\Gamma \approx 5$ MHz. The latter value corresponds to the hole observed on the absorption spectra and takes into account the laser linewidth (1 MHz), drift and other effects like power broadening.
 
In Fig. \ref{fig2}a, the absorption lineshape  does not change with pumping duration and the fraction of pumped ions is small.  The peak absorption coefficient measured at 0 GHz and normalized by its value without OP, i.e. at thermal equilibrium,  is therefore equivalent to $k_{4g}$.

%$\Gamma = 5$ MHz square region is pumped (taking account power broadening, laser drift and spectral diffusion) within the $\Gamma_0 = 500$ MHz broad Lorentzian line (observed in the 10 ppm sample \cite{Ortu:2018ig}), the A/B concentration ratio is : 
%\begin{equation}
%C = \frac{C_A}{C_B} \approx \frac{2}{\pi}\frac{\Gamma}{\Gamma_0} = 6.4 \times 10^{-3}.
%\end{equation}
%

\subsection{Optical coherence measurement}
For those measurements, a second AOM was added to enhance gating and avoid optical pumping during photon echo sequences. The beam was focused by a 100 mm focal length lens with a power of 7 mW. The photon echo was measured using a standard Hahn echo sequence ($\pi/2-\tau-\pi-\tau-\mathrm{echo}$) with durations of 1 and 2 $\mu$s for the $\pi/2$ and $\pi$ pulses. 
Due to the laser jitter, the echo amplitude significantly fluctuated for $\tau > 300\ \mu$s. To overcome this issue, for a given delay $\tau$, 50 successive echo sequences were recorded and only the strongest echo was kept. We checked that the echo sequences themselves did not cause optical pumping. Echo pulse power was also varied to look for instantaneous spectral diffusion, which was not observed.

\bibliography{Papers}

\onecolumngrid
\newpage

\section*{Coherence enhancement by optically induced electron spin polarization \\ in a rare-earth doped single crystal
 - Supplementary information}

\section{Optical pumping}

\subsection{Populations}
\label{Ratios}

\begin{figure}
\includegraphics[scale=0.8]{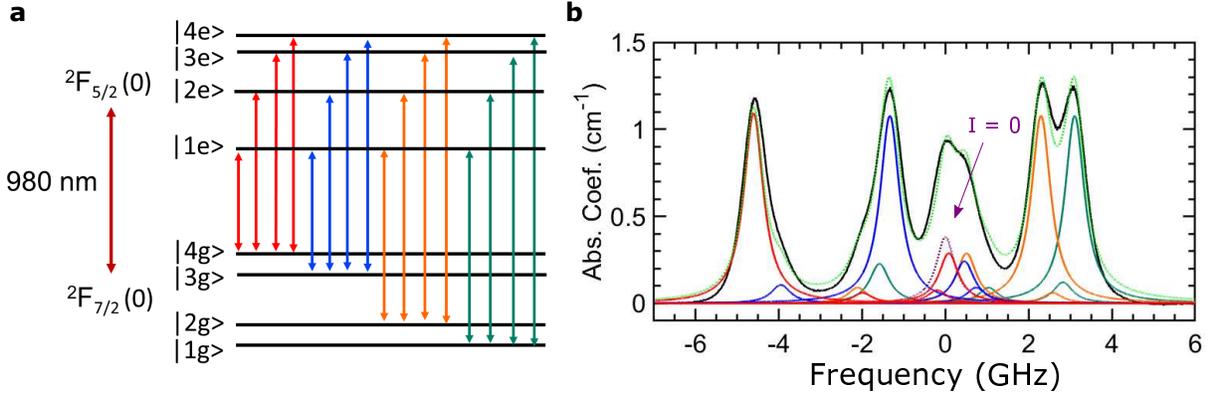}
\caption{\textbf{a.} Energy diagram of the hyperfine structure of \ybi\ in  \YSO\ in site 2 for \ybground(0) and \ybexcited(0). The different colors correspond to the optical transitions connecting the same ground state spin level. \textbf{b.} Absorption spectrum at 2 K without prior OP. The different optical transitions constituting the whole absorption spectrum are shown. Their color correspond to the arrow ones in the energy diagram. The residual absorption line from $I$ = 0 \yb\ isotopes is observed. The green dashed line represents the fitted curve $\alpha(\nu)$ defined in Eq. \eqref{popratio}.}
\label{deconvo_eq}
\end{figure}

Since we are using an isotopically purified sample, the main contribution to the absorption spectra are composed of the 16 different optical transitions belonging to the $S=1/2,I=1/2$ \ybi\ ions.  A weak contribution of the $S=1/2,I = 0$ nuclear spin isotopes ($\approx 6$\% of all \yb\ ions), is observed at 0 GHz, see Fig. \ref{deconvo_eq}a.
All spectra were fitted using the expression:
\begin{equation}
\alpha(\nu) = \alpha_0\sum_{i=1..4,j=1..4} k_{ig} \beta_{ij} g(\nu, \nu_{0,ij}, \Gamma_0) +\alpha_1 g(\nu, \nu_1, \Gamma_1). \label{popratio}
\end{equation}
in which the first term on the right hand side corresponds to \ybi\ and the second term to $I=0$ isotopes.
In Eq. \eqref{popratio},  $\alpha(\nu)$ and $\alpha_{0,1}$ are absorption coefficients, $\nu$ the frequency, $i$ ($j$) ground (excited) state labels and $\beta_{ij}$ the branching ratio between levels $i$ and $j$ (see Table \ref{branching_ratios} and \cite{Businger:2019vw}). $g(\nu,\nu_k,\Gamma_k)$ are area-normalized Lorentzian functions with full width at half maximum $\Gamma_{k}$ and center frequency $\nu_k$ ($\int g(\nu,\nu_k,\Gamma_k)d\nu=1$). Center frequencies were determined from hole burning experiments \cite{Tiranov:2018kx} and correspond to the scheme in Fig. \ref{deconvo_eq}. The ground state populations $k_{ig}$ verify $0\leq k_{ig} \leq 4$ and $\sum k_{ig} =4$. 
These are normalized to their thermal equilibrium values, which for our working temperatures are $k_{ig,eq}=1$, to a good approximation.

\begin{table}[h]
   \begin{tabular}{c|cccc}
              & $\ket{1_e}$ & $\ket{2_e}$ & $\ket{3_e}$ & $\ket{4_e}$\\
   \hline
    $\ket{1_g}$ & 0.15 & 0.06 & 0.08 & 0.71\\
	$\ket{2_g}$ & 0.06 & 0.19 & 0.71 & 0.04\\
	$\ket{3_g}$ & 0.07 & 0.71 & 0.16 & 0.06\\
	$\ket{4_g}$ & 0.72 & 0.04 & 0.05 & 0.19\\   
   \end{tabular}
\label{branching_ratios}
\caption{Branching ratios  for \ybi:\YSO\ in site 2 and 
 light polarized along $D_2$ axis using spectral holeburning technique \cite{Businger:2019vw}.}
\end{table}

The absorption spectrum recorded at 2 K without prior optical pumping (OP) was first used to determine $\Gamma_{0,1}$ and $\alpha_{0,1}$. In this case, the \ybi\ population is equally distributed into the four hyperfine ground states and $k_{ig}=k_{ig,eq}=1$. Best fit values were $\Gamma_0=572 \pm 40$ MHz, $\Gamma_1$ = 540 $\pm\ 100$ MHz, $\alpha_0 = $ 1.37 $\pm\ 0.16$ \wnm\ and $\alpha_1$ = 0.32 $\pm\ 0.08$ \wnm. Experimental and fitted spectra are shown in Fig. \ref{deconvo_eq}b. 

\begin{figure}
\includegraphics[scale=0.8]{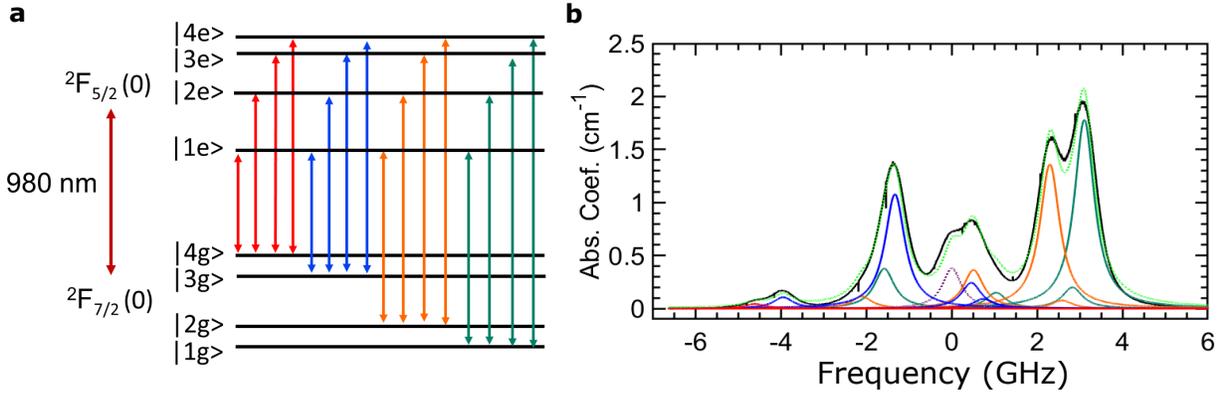}
\caption{\textbf{a,b} Same as in Fig. \ref{deconvo_eq}a,b but with the absorption spectrum recorded after 20 s of OP at -4.8 GHz (Fig. 1c of the main text). The fit gives $k_{1g}, k_{2g}, k_{3g}, k_{4g} = 1.67 \pm 0.30, 1.28 \pm 0.36, 1.01 \pm 0.12, 0.04 \pm 0.04$.}
\label{deconvo_after_OP}
\end{figure}

Spectra obtained after DEOP were fitted by varying $k_{ig}$ coefficients while keeping $\Gamma_{0,1}$ to the previous values. In some cases, adjustments of $\alpha_{0}$ ($\pm\ 6\ \%$) and $\alpha_{1}$ ($\pm\ 10\ \%$) were needed due to small changes in experimental conditions (light polarization, beam alignment with respect to sample). Fig. \ref{deconvo_after_OP}b displays the fit of the spectra in Fig. 1c of the main text resulting in $k_{1g}, k_{2g}, k_{3g}, k_{4g} = 1.67 \pm 0.30, 1.28 \pm 0.36, 1.01 \pm 0.12, 0.04 \pm 0.04$. 

%DEOP At 2 K and  thermal equilibrium, the \ybi\ population is equally distributed into the four lowest hyperfine states. The whole absorption spectrum can be well fitted to a sum of 17 Lorentzian curves having center frequencies corresponding to the energy diagram shown in figure 1b. in the main text. The lines belonging to \ybi\ have a 550 MHz linewidth and their intensity, for light polarized along $D_2$, is proportional to the branching ratios given in Table: 

%matrix $\boldsymbol{\gamma}_{II,D_2}$:
%\begin{equation}
%\boldsymbol{\alpha}_{II,D_2} = \alpha_{II,0}\times\boldsymbol{\gamma}_{II,D_2}
%\end{equation}
%with
%\begin{equation}
%\boldsymbol{\gamma}_{II,D_2} = \begin{pmatrix}
%0.14 & 0.07 & 0.05 & 0.74\\
%0.06 & 0.19 & 0.71 & 0.04\\
%0.08 & 0.68 & 0.17 & 0.07\\
%0.72 & 0.06 & 0.07 & 0.15\\
%\end{pmatrix}
%\end{equation}
%
%The best fit to the experimental spectrum was obtained by scaling the sum of all Lorentzian curves by a coefficient equals to 1.46 cm$^{-1}$. 
%Table 1 values were measured in the same way as in in \cite{Businger:2019vw} (Table1, SI), but in different samples, which explains small differences. The 17$^{\mathrm{th}}$ Lorentzian curve, belonging to the $I=0$ isotopes, placed at 0 GHz frequency detuning in the spectrum in figure S\ref{deconvo_eq}b , has a linewidth of 500 MHz and a peak absorption coefficient of 0.44 cm$^{-1}$.\\

%After the optical pumping, we can use the absorption spectrum to get the population ratios $k$ = ($k_1$, $k_2$, $k_3$, $k_4$), which correspond to the ratio of the population in each of the ground states with respect with their value at the equilibrium. In another way, at the thermal equilibrium we have : $k_{eq}$ = (1, 1, 1, 1).
%The population ratios $k$ are fitted to the absorption spectrum using :
%\begin{equation}
%\boldsymbol{\alpha}_{II,D_2} = \alpha_{II,0}\times(\mathbf{k}\cdot\boldsymbol{\gamma}_{II,D_2})
%\end{equation}
%with
%\begin{equation}
%\mathbf{k} = \begin{pmatrix}
%k_1 & k_1 & k_1 & k_1\\
%k_2 & k_2 & k_2 & k_2\\
%k_3 & k_3 & k_3 & k_3\\
%k_4 & k_4 & k_4 & k_4\\
%\end{pmatrix}
%\end{equation}
%The population ratios determined according to the orange spectrum in the figure 1c. are $k$ = (1.62, 1.28, 1.06, 0.04), see figure \ref{deconvo_after_OP}. %\SW{According to the experimental fluctuations and the post-treatment of the spectra, we estimate the standard error for those coefficients to be less than 5$\%$}. 

\section{DEOP Modeling}

\subsection{Qualitative mechanism}

\begin{figure}
\hspace{0mm}{
\includegraphics[scale=0.85]{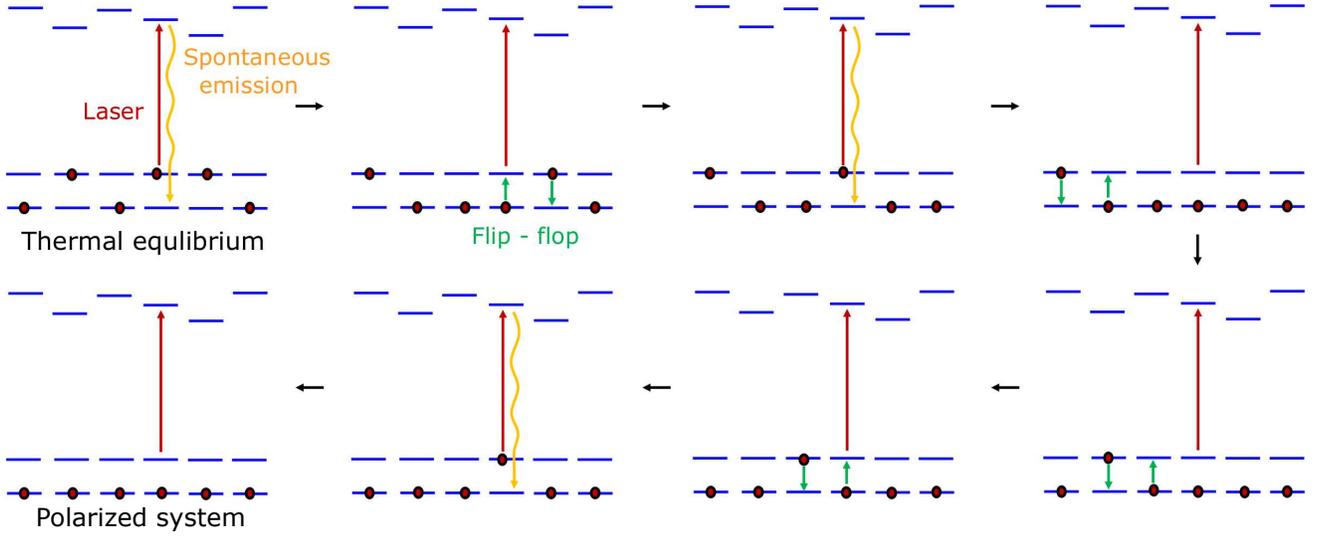}}
\caption{Qualitative mechanism for DEOP.}
\label{cartoon_diffusion}
\end{figure}

The cartoon shown in Fig. \ref{cartoon_diffusion} explains the mechanism of Diffusion Enhanced Optical Pumping using a simplified system. Here, the ground state is composed of two spin levels and we consider only one optically excited state. Optical pumping (OP) is applied along the transition connecting the higher ground state spin level to the excited state. Ions in the excited state can relax  towards the lower spin state, which results in spin polarization. However, since the optical transition is inhomogeneously broaden, the laser is resonant only with a subgroup of all  ions.   Relaxation of individual spins by interaction with the lattice  is assumed to be very slow.
%The ions next to each other represent the ions that are close enough and have a close enough spin transition energy to have a non-negligible flip-flop rate. 
At thermal equilibrium, both ground state spin levels are equally populated. DEOP can be described in the following way: an optically pumped spin initially in the higher spin state is transferred to the lower state. This spin can flip-flop with a non-pumped neighbor in the higher spin state. The pumped spin  goes back to the higher state, while the neighbor goes to the lower state. If that happens, the pumped spin is transferred  again to the lower state, resulting in two spins in the lower state.
The neighboring spin can further  flip-flop with another non-pumped spin in the higher state, after which the previous sequence can repeat, eventually leading to three spins in the lower state. In this way, the whole system of pumped and non-pumped spins can be completely polarized.
Qualitatively this requires that the optical pump rate is higher than the flip-flop rate, such that the optically pumped spin spends most of their time in their lower state, while the flip-flop rate should be higher than the spin lattice relaxation rate.

%Because of flip-flops between non-pumped spins, Step by step, the spin state will diffuse until the system reaches a polarized steady-state.

\subsection{Rate equation modeling of DEOP}
\label{DEOP}

\begin{figure}
\includegraphics[scale=0.65]{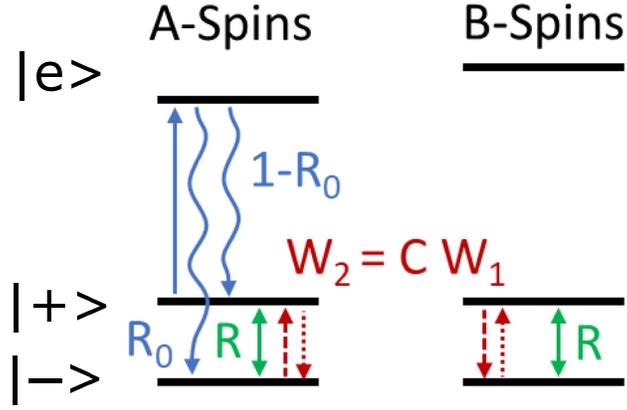}
\caption{Spin 1/2 model scheme.}
\label{spin_diffusion}
\end{figure}

Here, we use the same energy level scheme as described above for $^{171}$\yb{} spins. Therefore, instead of the 4 levels resulting from the low-symmetry anisotropic hyperfine interaction, we consider a $S$ = 1/2 spin system. The A-spins are optically pumped and assumed to be non-interacting with each other because of their low concentration. This is justified because the pump laser is much narrower than the optical inhomogeneous broadening. However, they interact with the non-pumped B-spins via magnetic dipole-dipole interactions, see Fig. \ref{spin_diffusion}. For A-spins, the optical pumping is resonant with the transition connecting the upper ground state level and the excited state. The effective pumping rate to the lower level is noted $R_0$. It takes into account the optical excitation rate $R_L$ from $\ket{+}$ to $\ket{e}$, the excited state radiative population lifetime $T_{1,o}$ and the branching ratio $\beta$ for  $\ket{e}\leftrightarrow\ket{-}$ transition. In our experimental conditions, $R_L\gg 1/T_{1,o}$ and $R_0 = \beta/T_{1,o}$. The population in the upper and lower ground states of A- and B-spins can also relax to each other  through spin-lattice relaxation (SLR) at the same rate rate $R$. 
%If a $\Gamma = 5$ MHz square region is pumped (which takes account power broadening, laser drift and spectral diffusion) within the $\Gamma_0 = 550$ MHz broad Lorentzian line, the A/B concentration ratio is : 
%\begin{equation}
%C = \frac{C_A}{C_B} \approx \frac{2}{\pi}\frac{\Gamma}{\Gamma_0} = 5.8 \times 10^{-3}.
%\end{equation}
Although each A-spin has presumably a different environment in terms of distances and directions of neighboring B-spins, we do not take it into account and consider an average probability $p_A^{+,-}$ for A-spins to be  in the upper or lower state.

The B-spins are not pumped, and, since they are in higher concentration, they interact with each other through resonant spin flip-flops mediated by the magnetic dipole-dipole interaction. This interaction is assumed to be fast enough on the time scale of the experiment so that all B-spins  have the same probability $p_B^{+,-}$ of occupying the upper or lower states. 
%This can be also justified in the case of a slow enough SLR relaxation, because flip-flops among B-spins will not change the average level populations, that we can identify to  $p_B^{+,-}$. 
%Only A-B flip-flops will change $p_B^{+,-}$, through A-spins optical pumping.
%
The rate equation for an A-spin i is:
\begin{equation}
\frac{dp_{A,i}^{+}}{dt} = - (R_0 + R)p_{A,i}^{+}+R(1-p_{A,i}^{+})-\sum_{j,B} \left [W_{ij}p_{A,i}^{+}(1-p_{B,j}^{+})-W_{ij}(1-p_{A,i}^{+})p_{B,j}^{+} \right ],
\label{rate1}
\end{equation}
where $W_{ij}$ is the flip-flop rate between A-spin $i$ and B-spin $j$. We used $p_{X,i}^{-} = 1-p_{Xi}^{+}$. According to the above assumptions, subscripts $i$ and $j$ can be dropped for probabilities and $\sum_{j,B} W_{ij}$ does not depend on $i$. Eq. \eqref{rate1} is then written as:
\begin{eqnarray}
\frac{dp_{A}^{+}}{dt} &=& - (R_0 + R) p_{A}^{+}+R(1-p_{A}^{+})-\left [\sum_{j,B} W_{ij}\right ] \left [p_{A}^{+}(1-p_{B}^{+})-(1-p_{A}^{+})p_{B}^{+} \right ] \nonumber \\ 
                                 &=&  - (R_0 + R) p_{A}^{+}+R(1-p_{A}^{+})-W_{1}\left [p_{A}^{+}(1-p_{B}^{+})-(1-p_{A}^{+})p_{B}^{+} \right ] \nonumber \\
                                 &=& - (R_0 + R) p_{A}^{+}+R(1-p_{A}^{+})-W_{1}\left [p_{A}^{+}-p_{B}^{+} \right ].
\label{rate2}
\end{eqnarray}

Similarly we have for B-spins:
\begin{eqnarray}
\frac{dp_{B}^{+}}{dt} &=& - R p_{B}^{+}+R(1-p_{B}^{+})-\left [\sum_{i,A} W_{ij}\right ] \left [p_{B}^{+}(1-p_{A}^{+})-(1-p_{B}^{+})p_{A}^{+} \right ] \nonumber \\
                                &=& - R p_{B}^{+}+R(1-p_{B}^{+})-W_ 2 \left [p_{B}^{+}(1-p_{A}^{+})-(1-p_{B}^{+})p_{A}^{+} \right ] \nonumber \\
                                &=&  - R p_{B}^{+}+R(1-p_{B}^{+})-W_ 2 \left [p_{B}^{+}-p_{A}^{+} \right ].
\label{rate3}
\end{eqnarray}
Since $W_1$ and $W_2$ are summed respectively on the B- and A-spins, we have $C = W_2/W_1$, where $C$ is the fraction of pumped spins.

The  steady state solution of \eqref{rate2} and \eqref{rate3} is obtained by setting the time derivatives on the left hand side to zero:
\begin{eqnarray}
0 &=& - (R_0+R) p_{A}^{+,\infty}+R(1-p_{A}^{+,\infty})-W_{1}\left [p_{A}^{+,\infty}-p_{B}^{+,\infty} \right ] \\
0 &=&- R p_{B}^{+,\infty}+R(1-p_{B}^{+,\infty})-W_ 2 \left [p_{B}^{+,\infty}-p_{A}^{+,\infty} \right ],
\end{eqnarray}
which gives:
\begin{eqnarray}
p_{A}^{+,\infty} &=& \frac{R (2R+W_1+W_2 )}{2R (R_o+W_1 ) + W_2(R_o+2R)+4R^2}  \\
p_{B}^{+,\infty} &=& \frac{R (R_o+2R+W_1+W_2 )}{2R (R_o+W_1 ) + W_2(R_o+2R)+4R^2}.
\end{eqnarray}

We further assume $R_o \gg R,W_2$ and $W_1\gg R,W_2$, because in our system the SLR $R$ rate is small, the optical pumping rate $R_0$ is strong, and only a small fraction of ions is pumped ($C=W_2/W_1 \ll 1$), as discussed in the main text. We introduce 
$\beta_o = R/(R_o+R)\approx R/R_0$ and 
$\beta_{ff}= W_1/(R_o+W_1)$ and
obtain:
\begin{eqnarray}
p_{A}^{+,\infty} &\approx&\frac{W_1}{R_o+W_1} \left ( 2+\frac{R_oW_1}{R(R_o+W_1)} C \right )^{-1}\\
&\approx& \frac{\beta_{ff} }{ 2+C\beta_{ff}/\beta_0  } \label{eq:paf}\\
p_{B}^{+,\infty} &\approx& \left ( 2+\frac{R_oW_1}{R(R_o+W_1)} C \right )^{-1} \label{eq:pbf} \\
&\approx& \frac{1}{ 2+C \beta_{ff}/\beta_o}.
\end{eqnarray}

%The steady state population of B-spins lowest state deacreases  The DEOP effect steady state is therefore increased when $C$, and $\beta_{ff}$ increase and $\beta_o$ decreases. This means that it increases with stronger optical pumping, slower SLR, larger fraction of pumped ions and stronger flip-flops. 

The rates at which the steady states are  reached can also be obtained from the system of equations \eqref{rate2} and \eqref{rate3}. Under the assumption $R,W_2\ll R_o,W_1$, there is a fast component with a rate $R_o+R$ and a slow one, called the polarization rate in the main text,  with a rate:
\begin{equation}
R_P =  R(2+C\beta_{ff}/\beta_o).
\end{equation}
$R_P$ is the rate that is determined from the experiments of Fig. 2b, main text.

%We see that the DEOP characteristic time decreases with $C,\beta_{ff}$ and increases with $\beta_o$. We note that $R_P$ is simply proportional to $C$. In our experiment, considering the branching ratio of the considered optical transition is 75\%, we can estimate the rate $R_0\approx (1-0.75)/T_{o} = 192$ s$^{-1}$. $T_{o}$ represents the radiative lifetime of the excited state.
%
%If we assume that $R_o\gg W_1$, which is only partially verified experimentally (see below),  we can reduce $R\beta_{ff}/\beta_o$ to $W_1$ and $\beta_{ff}/\beta_o$ to $W_1/R$. Experimental data and fits using equations  \eqref{eq:paf} and \eqref{eq:pbf} are shown in Fig. 2c in the main text. \SW{(@Philippe : I changed a little bit the values of C considering a Lorentzian of a width of 550 MHz. But the fit is much worse...)} Those fits give $R = 1.8 \times 10^{-2}\pm 1 \times 10^{-2}$ $s^{-1}$ and $W_1=58 \pm 10 $ $s^{-1}$. The value of $R$ is consistent with the value determined experimentally of $1/72 = 1.38 \times 10^{-2}$ $s^{-1}$.
%
\subsection{Spin flip-flops}

For simplicity we assume the Zeeman $g$- and hyperfine $A$-tensors are diagonal in the same basis and have anisotropic form
$g_x \neq g_y \neq g_z$ and $A_x \neq A_y \neq A_z$. This assumption is well justified for certain solid-state systems (for example \ybiso{} crystal \cite{Tiranov:2018kx}). In this case, at zero magnetic field, the wavefunctions are given only by the hyperfine tensor, which makes all the levels to be non-degenerate: 
$$\ket{1} =  (\ket{\uparrow \Uparrow} - \ket{\downarrow \Downarrow})/\sqrt{2},  \text{ } \ket{2} = (\ket{\uparrow \Uparrow} + \ket{\downarrow \Downarrow})/\sqrt{2},$$
$$\ket{3} =  (\ket{\uparrow \Downarrow} - \ket{\downarrow \Uparrow})/\sqrt{2}, \text{ } \ket{4} =  (\ket{\uparrow \Downarrow} + \ket{\downarrow \Uparrow})/\sqrt{2}. $$

All the spin transitions, in this situation,  are connected purely by $S_x$ ($ \ket{1} \leftrightarrow \ket{4}$ and $\ket{2} \leftrightarrow \ket{3}$), $S_y$ ($ \ket{1} \leftrightarrow \ket{3}$ and $\ket{2} \leftrightarrow \ket{4}$) or $S_z$ ($ \ket{1} \leftrightarrow \ket{2}$ and $\ket{3} \leftrightarrow \ket{4}$) spin 1/2 operators.  This strongly simplifies the expression for the dipole-dipole interaction $H_{dd} $  that will contain only corresponding operators. In this case the flip-flop rate estimated by Fermi golden rule $|\bra{i,f} H_{dd} \ket{f,i}|^2$ will be proportional to the corresponding element of the $g$-tensor \cite{Cruzeiro:2017hp,Car:2018to}:
\begin{center}
for $ \ket{1} \leftrightarrow \ket{2}$ and $\ket{3} \leftrightarrow \ket{4}$:  $\propto g_z^4$

for $ \ket{1} \leftrightarrow \ket{3}$ and $\ket{2} \leftrightarrow \ket{4}$:  ${\propto g_y^4},$

for $ \ket{1} \leftrightarrow \ket{4}$ and $\ket{2} \leftrightarrow \ket{3}$:  ${\propto g_x^4}.$
\end{center}

The anisotropy of the $g$-tensor can lead to dramatically different relaxation times for different transitions. In the case of  \ybi\ in site 2 of \YSO, $g_z$ = 6.06, $g_y$ = 1.5, $g_x$ = 0.13 \cite{Welinski:2016gi}. This predicts a few orders of magnitude variation for different transitions, flip-flops within the $\ket{1} \leftrightarrow \ket{2}$ and $\ket{3} \leftrightarrow \ket{4}$ pairs of levels being much faster than all the other ones. 

\subsection{SLR modeling}

The recovery time $R_c$ of the $\ket{4g} \leftrightarrow \ket{1e}$ optical line after the OP has been stopped corresponds to spin lattice relaxations. Indeed spin flip-flops do not change overall populations in the case of no optical pumping. $R_c$ has been measured for different temperatures and its variations are shown on Fig 2d in the main text. Those variations can be modeled by considering the one-phonon direct process and two-phonons processes \cite{Kiel:1967ch}:
\begin{equation}
R_c = \alpha_D\mathrm{coth}(\frac{h\nu_{eff}}{2k_BT}) + \alpha_{R}\int_0^{\frac{\pi}{2}}\frac{q^8e^{-\frac{\theta_D}{T}\mathrm{sin}q}\mathrm{d}q}{(1-e^{-\frac{\theta_D}{T}\mathrm{sin}q})^2(\theta_E^2-\theta_D^2\mathrm{sin}^2q)^2}
\end{equation}
In this equation, the same parameters than in a previous study \cite{Lim:2018hs}, $\theta_D$ = 100 K and $\theta_E$ = 337 K, have been used for the two-phonon part. 
%At zero magnetic field, the spin structure in composed of four non degenerate levels. Therefore, we can expect a direct process happening between those levels \cite{Larson:1966jy}. 
The direct process term uses  the average splitting between $\ket{4g}$ and the three other ground state spin levels,  $\nu_{eff}$ = 2.06 GHz. The fitted coefficients are $\alpha_D = 3.9\times10^{-4}$ s$^{-1}$ and $\alpha_{R} = 0.9\times10^{18}$ s$^{-1}$.K$^4$. The latter value is  reasonably close to the one determined in \cite{Lim:2018hs}.

\subsection{Narrow hole decays}

\begin{figure}
\includegraphics[scale=0.6]{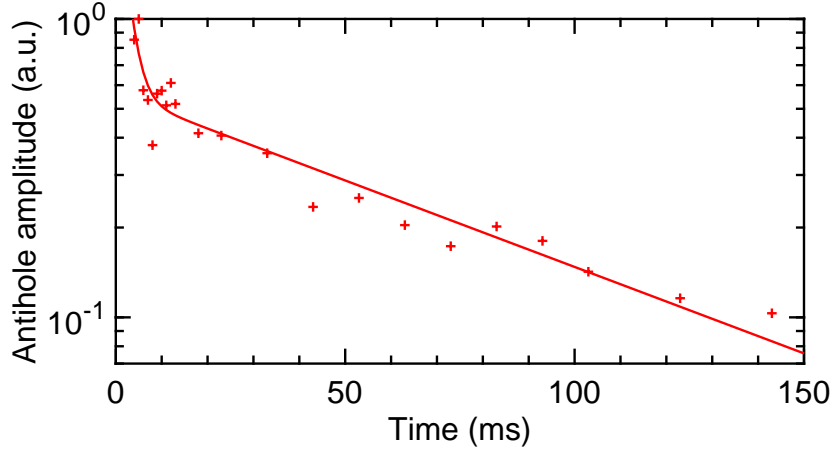}
\caption{Decay of an anti-hole  in the $\ket{4g} \leftrightarrow \ket{1e}$ line (crosses) and two-exponential fit (solid line).}
\label{SHB_measurement}
\end{figure}

In order to compare the $W_1$ value extracted from the fit in Fig. 2c (main text) to experimental flip-flop rates, we investigated the dynamics of a narrow anti-hole in the $\ket{4g} \leftrightarrow \ket{1e}$ transition. It was obtained by  burning a hole close to the center of the $\ket{2g} \leftrightarrow \ket{4e}$ transition at +2.5 GHz (see Fig. 1c, main text) for 10 ms, a duration short enough to  avoid DEOP. The anti-hole height was measured at varying delays after the hole burning. Each measurement was preceded by an initialization sequence of 50 pulses scanned over 10 GHz  to prevent accumulating populations. Fig. \ref{SHB_measurement} shows the experimental data together with a two-exponential fit, giving rates of 500 and 13 s$^{-1}$. The larger rate is attributed to the fast $\ket{4g} \leftrightarrow \ket{3g}$  flip-flops and the other one, which correspond to $W_1$ in the rate equation model, to the intermediate $\ket{4g} \leftrightarrow \ket{2g}$  flip-flops.

%
%  The sequence began with an initialization pulse (see the corresponding section below) followed by an OP pulse of variable duration with a laser set at -1.5 GHz with respect to the center of the absorption spectrum, see figure \ref{SHB_measurement}b. This frequency is resonant with a part of the transitions $\ket{3g}-\ket{2e}$ and $\ket{1g}-\ket{1e}$. The absorption spectra corresponding after the different OP duration (0s, 1s and 10s) are shown on \ref{SHB_measurement}b. 100 ms after the end of the OP pulse, a 10 ms pulse ("burning pulse") is sent with the laser frequency set at +2.5 GHz, resonant with a part of the transitions $\ket{1g}-\ket{3e}$, $\ket{1g}-\ket{4e}$, $\ket{2g}-\ket{3e}$ and $\ket{2g}-\ket{4e}$. This delay allows the piezo-electric ceramic controlling the laser frequency to stabilize, and therefore prevents frequency drifts during the burning pulse. Figure \ref{SHB_measurement}c. shows to the absorption spectrum measured 5 ms after the burning pulse and a prior OP of 1 s. We can observe the narrow hole at the position of the laser (+2.5 GHz), shown by a green arrow. We also observe several antiholes corresponding to the different classes of ions pumped into the $\ket{4g}-\ket{1e}$ transition. One of them is selected (labeled with a black arrow), and its amplitude is measured with respect with delay time after the burning pulse.
%
%Short pumping duration was used in order to avoid DEOP.

%

\section{Optical coherence}

\subsection{Populations}

\begin{figure}
\includegraphics[scale=0.6]{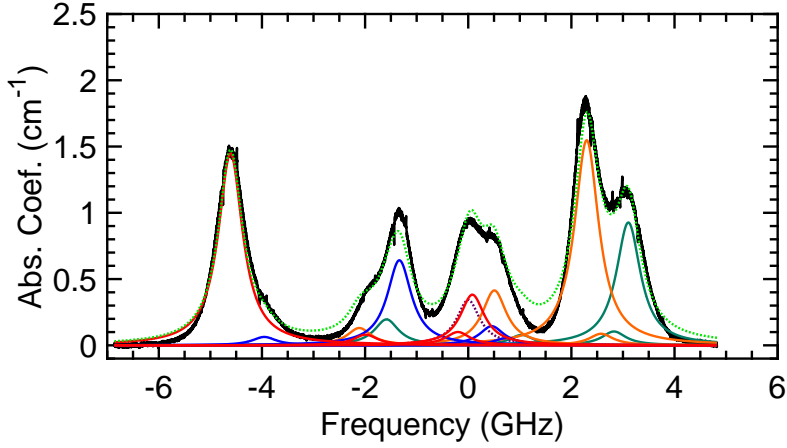}
\caption{Same as in Fig. \ref{deconvo_eq}a,b  but with the absorption spectrum recorded after 0.5 s OP at -1.44 GHz.}
\label{spectrum_1g3g1e2e_0o5s}
\end{figure}

\begin{figure}
\includegraphics[scale=0.6]{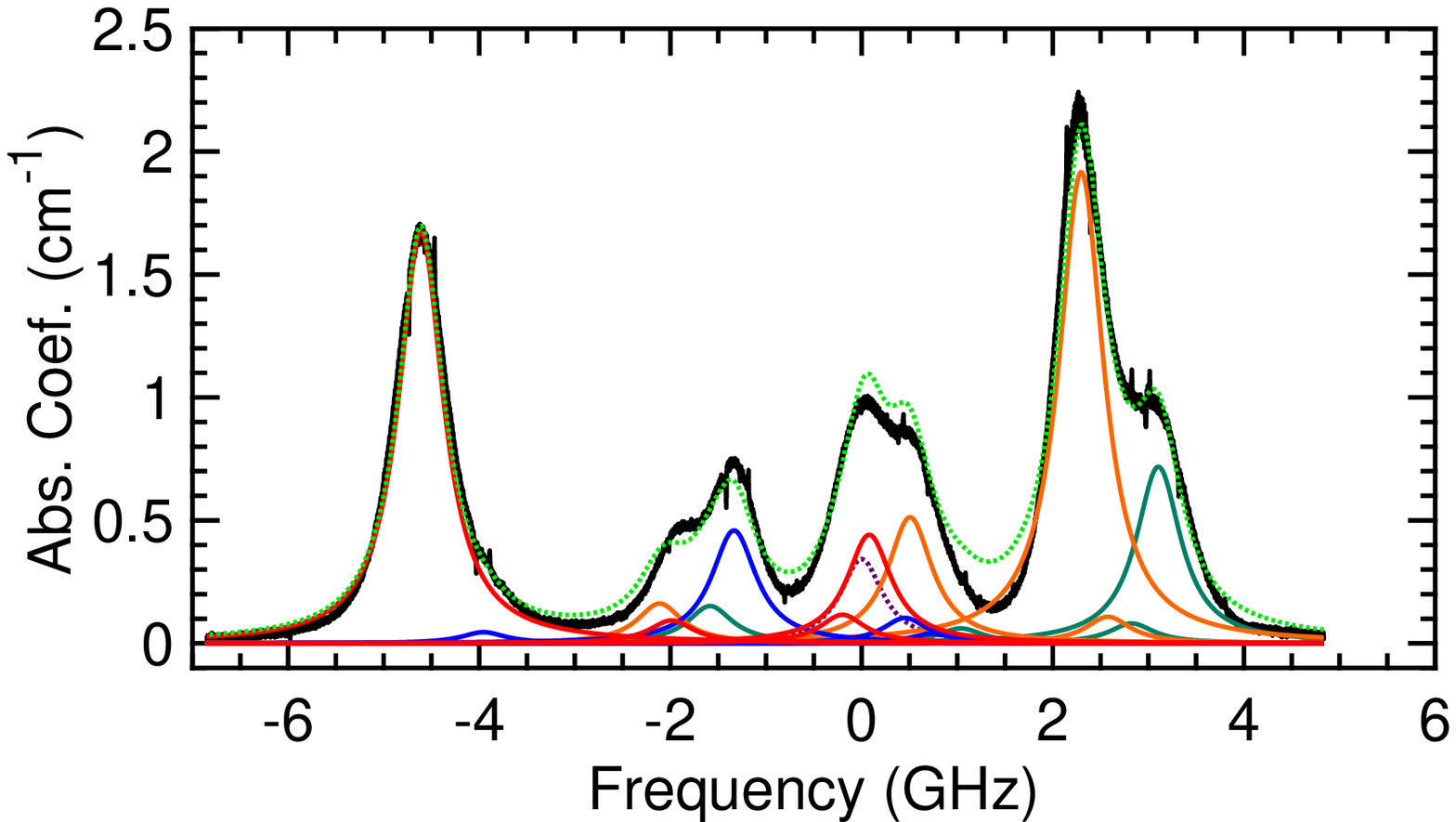}
\caption{Same as Fig. \ref{spectrum_1g3g1e2e_0o5s}, but after 1 s OP at -1.44 GHz.}
\label{spectrum_1g3g1e2e_1s}
\end{figure}

\begin{figure}
\includegraphics[scale=0.6]{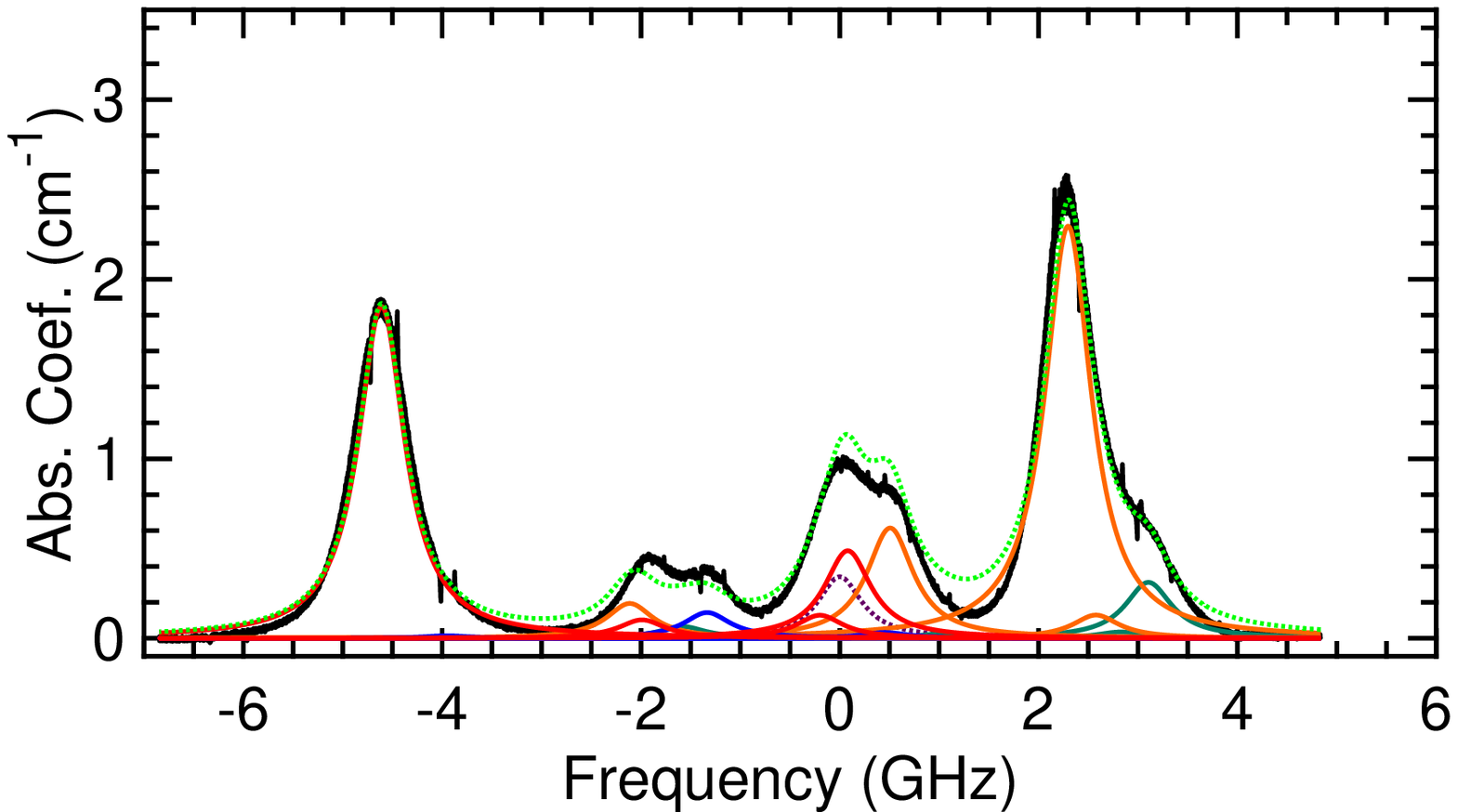}
\caption{Same as Fig. \ref{spectrum_1g3g1e2e_0o5s}, but after 2 s OP at -1.44 GHz.}
\label{spectrum_1g3g1e2e_2s}
\end{figure}

\begin{figure}
\includegraphics[scale=0.6]{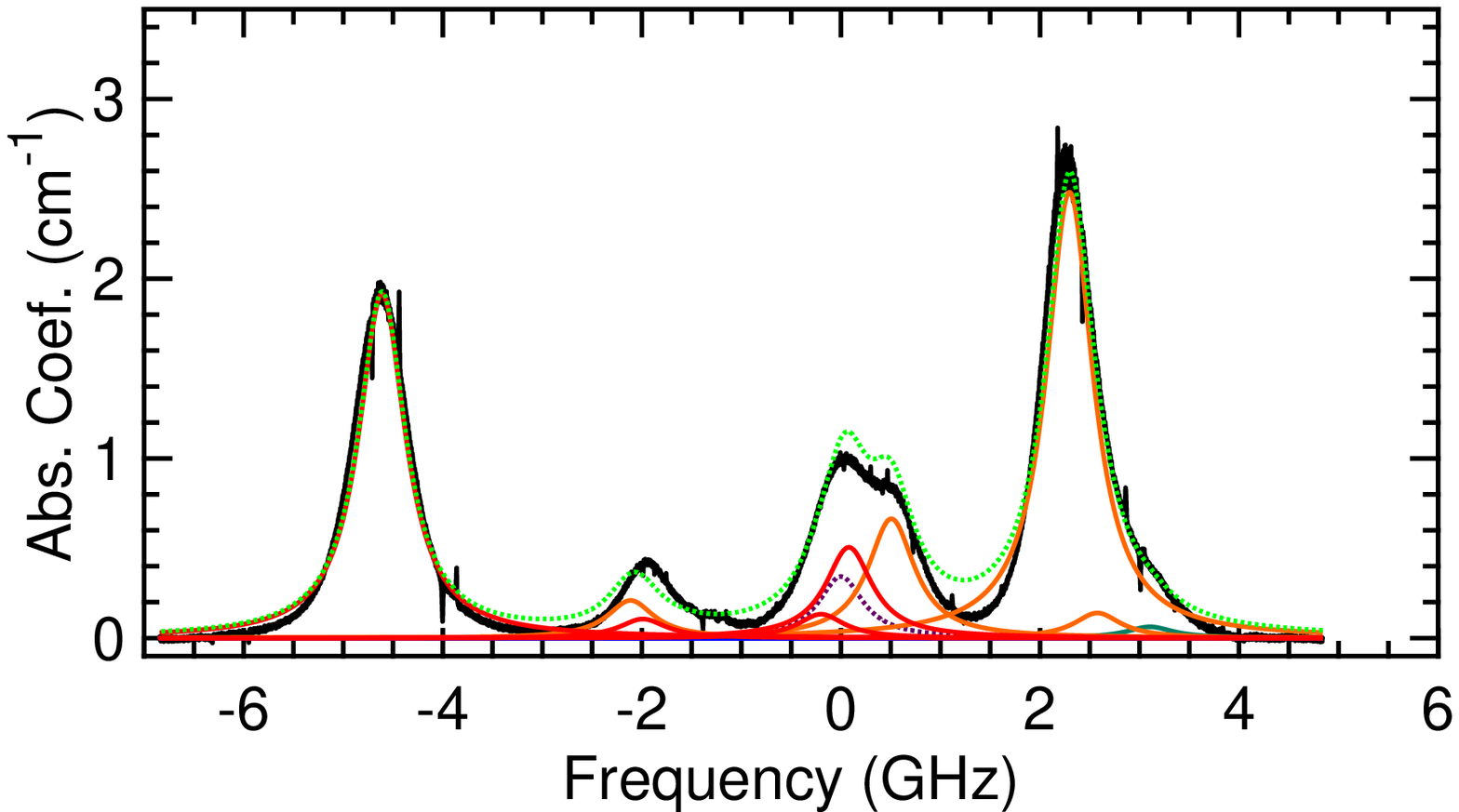}
\caption{Same as Fig. \ref{spectrum_1g3g1e2e_0o5s}, but after 5 s OP at -1.44 GHz.}
\label{spectrum_1g3g1e2e_5s}
\end{figure}

\begin{figure}
\includegraphics[scale=0.6]{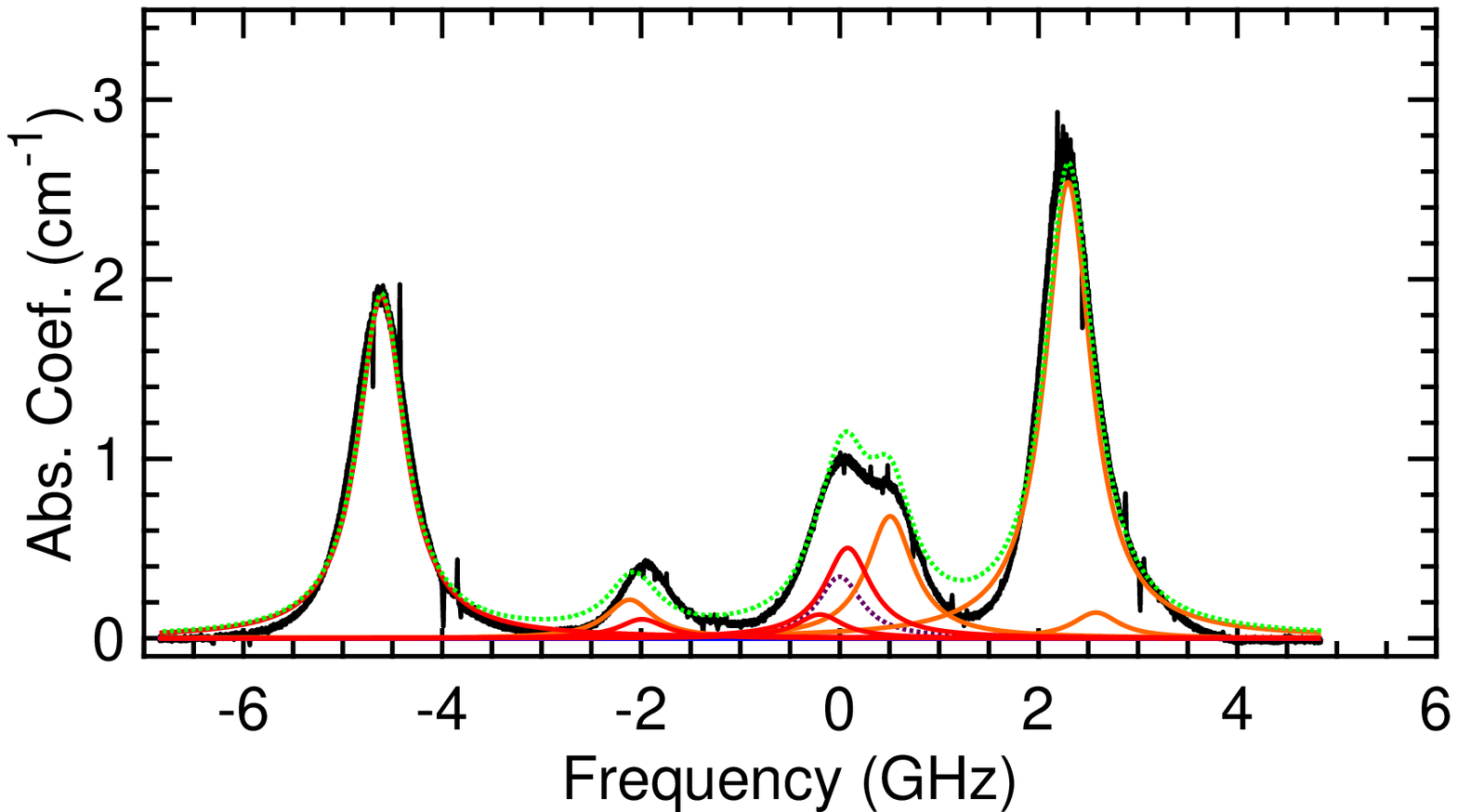}
\caption{Same as Fig. \ref{spectrum_1g3g1e2e_0o5s}, but after 10 s OP at -1.44 GHz GHz.}
\label{spectrum_1g3g1e2e_10s}
\end{figure}

\begin{figure}
\includegraphics[scale=0.6]{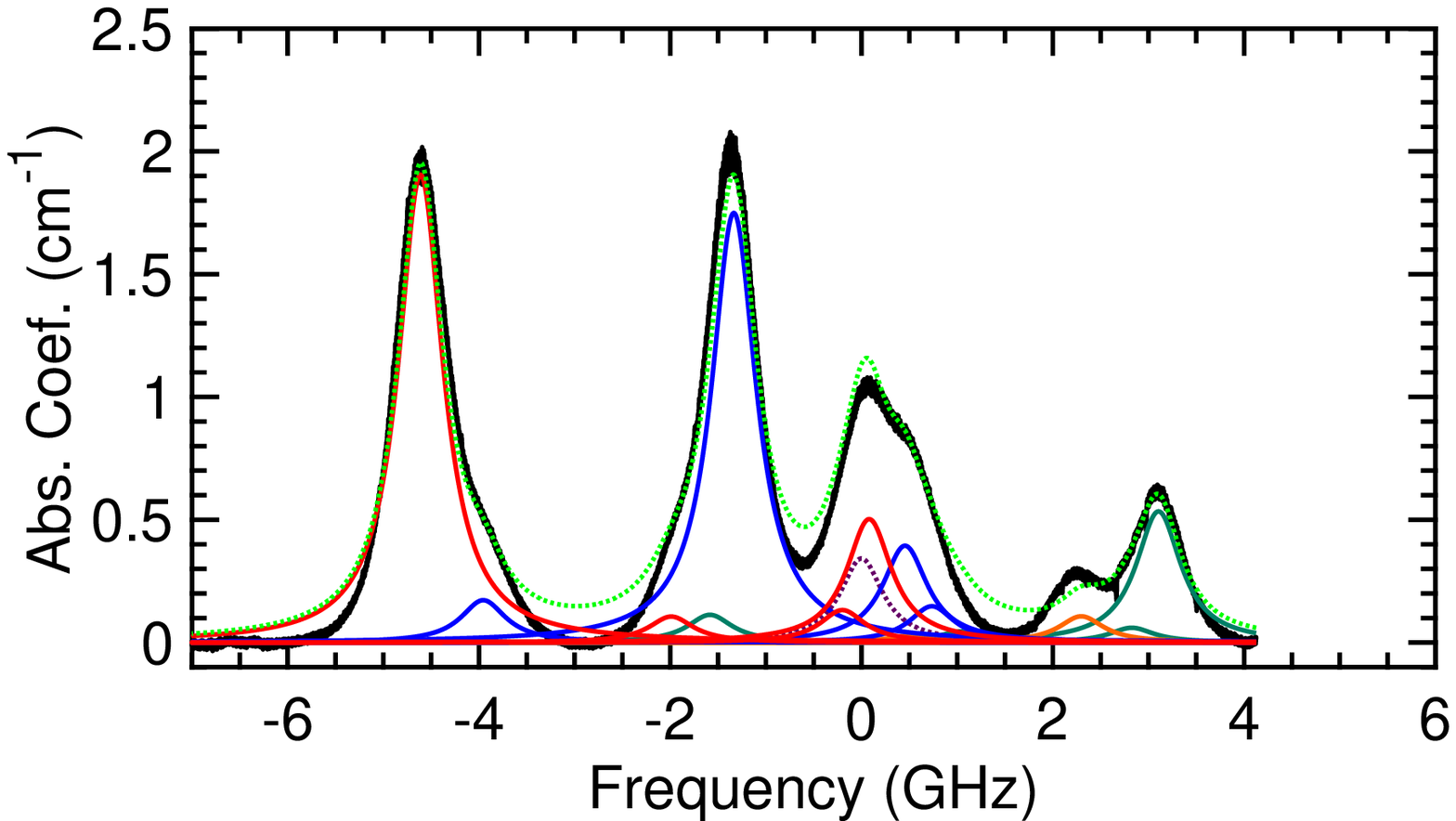}
\caption{Same as Fig. \ref{spectrum_1g3g1e2e_0o5s}, but after 1 s OP at +2.67 GHz.}
\label{spectrum_1g2g3e4e_1s}
\end{figure}

\begin{figure}
\includegraphics[scale=0.6]{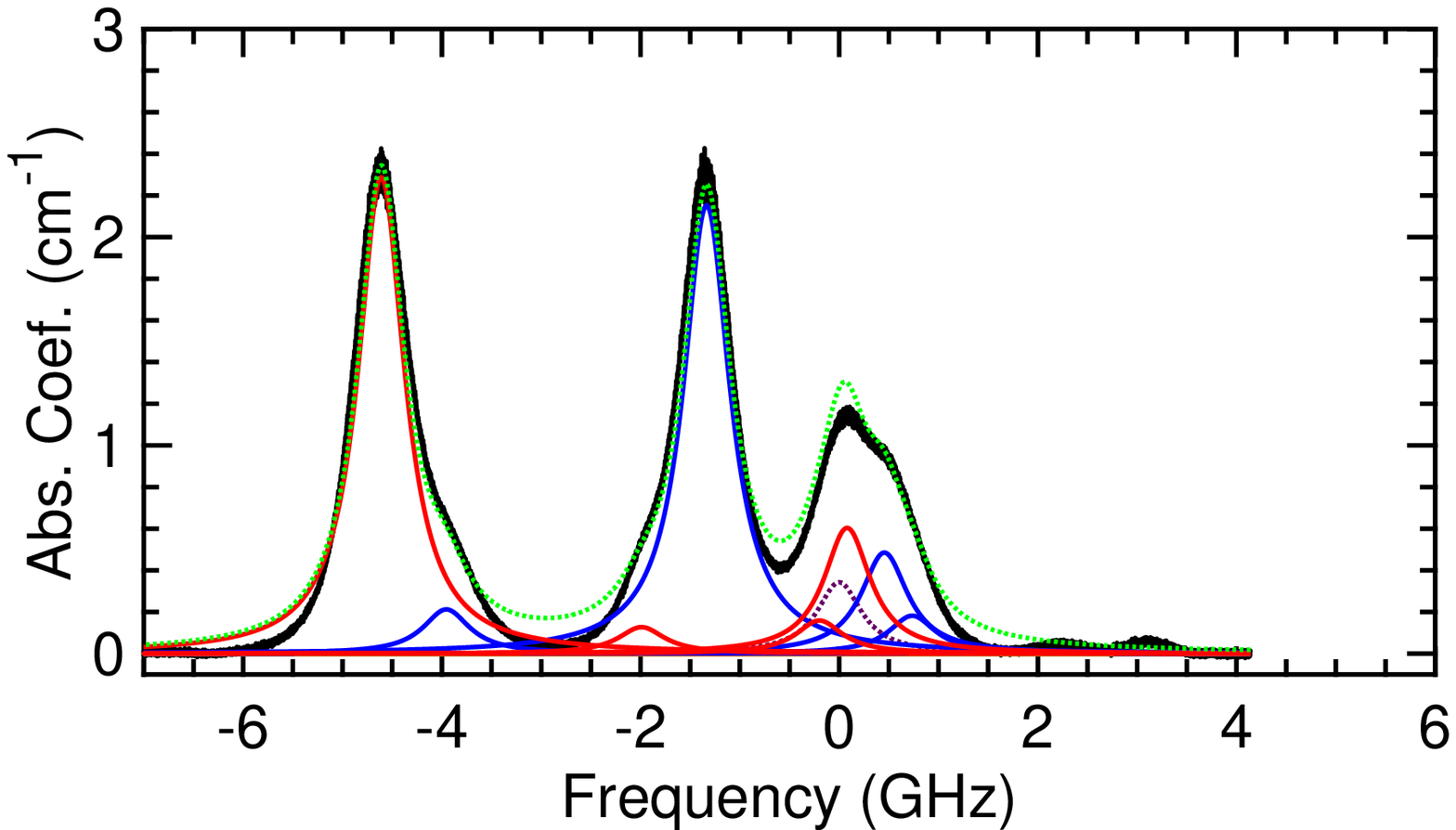}
\caption{Same as Fig. \ref{spectrum_1g3g1e2e_0o5s}, but after 10 s OP at +2.67 GHz.}
\label{spectrum_1g2g3e4e_10s}
\end{figure}

For echo measurements under DEOP (Fig. 3, main text), normalized populations ($k_{1g}$,$k_{2g}$,$k_{3g}$,$k_{4g}$) were determined from absorption spectra, as described in section \ref{Ratios}. The spectra and fits are shown in figures \ref{spectrum_1g3g1e2e_0o5s} to \ref{spectrum_1g2g3e4e_10s}, and the corresponding $k_{ig}$ values  presented in Table \ref{population_ratios_table}.

\begin{table}
\begin{tabular}{|c|c||c|c|c|c|}
\hline
OP condition & Absorption spectrum & $k_1$& $k_2$& $k_3$& $k_4$ \\
\hline
\hline
No prior OP & Figure \ref{deconvo_eq} & \textbf{1} & \textbf{1} & \textbf{1} & \textbf{1} \\
\hline
0.5 s OP laser at  -1.44 GHz& Figure \ref{spectrum_1g3g1e2e_0o5s} & \textbf{0.82} $\pm\ 0.06$ &   \textbf{1.36} $\pm\ 0.08$& \textbf{0.56} $\pm\ 0.08$& \textbf{1.26} $\pm\ 0.06$\\
\hline
1 s OP laser at  -1.44 GHz& Figure \ref{spectrum_1g3g1e2e_1s} & \textbf{0.61} $\pm\ 0.06$& \textbf{1.62} $\pm\ 0.08$&   \textbf{0.39} $\pm\ 0.06$&  \textbf{1.39} $\pm\ 0.06$\\
\hline
2 s OP laser at  -1.44 GHz& Figure \ref{spectrum_1g3g1e2e_2s} & \textbf{0.27} $\pm\ 0.04$& \textbf{2.01} $\pm\ 0.08$&   \textbf{0.13} $\pm\ 0.12$& \textbf{1.59} $\pm\ 0.06$\\
\hline
5 s OP laser at  -1.44 GHz& Figure \ref{spectrum_1g3g1e2e_5s} & \textbf{0.06} $\pm\ 0.04$ & \textbf{2.24} $\pm\ 0.08$ &   \textbf{0.05} $\pm\ 0.05$ & \textbf{1.70} $\pm\ 0.08$ \\
\hline
10 s OP laser at  -1.44 GHz& Figure \ref{spectrum_1g3g1e2e_10s} & \textbf{0.02} $\pm\ 0.02$ &   \textbf{2.30} $\pm\ 0.12$ &   \textbf{0.02} $\pm\ 0.02$ & \textbf{1.70} $\pm\ 0.06$ \\
\hline
\hline
1 s OP laser at +2.67 GHz& Figure \ref{spectrum_1g2g3e4e_1s} & \textbf{0.50} $\pm\ 0.02$ & \textbf{0.10} $\pm\ 0.02$ &   \textbf{1.64} $\pm\ 0.08$ & \textbf{1.76} $\pm\ 0.04$ \\
\hline
10 s OP laser at +2.67 GHz& Figure \ref{spectrum_1g2g3e4e_10s} & \textbf{0.02} $\pm \ 0.02$ &  \textbf{0.02} $\pm \ 0.02$ &  \textbf{1.95} $\pm\ 0.10$ & \textbf{2.05} $\pm\ 0.08$ \\
\hline
\end{tabular}
\caption{Normalized ground state populations corresponding to   the different OP conditions.}
\label{population_ratios_table}
\end{table}

\subsection{Spin coherence}

We performed another set of measurements to investigate the effect of the optical pumping on the spin coherence at 3~K. For this, we measured the spin coherence time $T_{2,s}$ through optical detection of a spin echo in a Hahn sequence using Raman heterodyne scattering (RHS), see \cite{Ortu:2018ig}. All spin echo measurements were carried out on the $\ket{4g}\leftrightarrow\ket{3g}$ transition (655 MHz) of site 2 by varying total population in $\ket{3g}$ and $\ket{4g}$ states for different optical pumping conditions. 

\begin{figure}
	\includegraphics[width=0.7\linewidth]{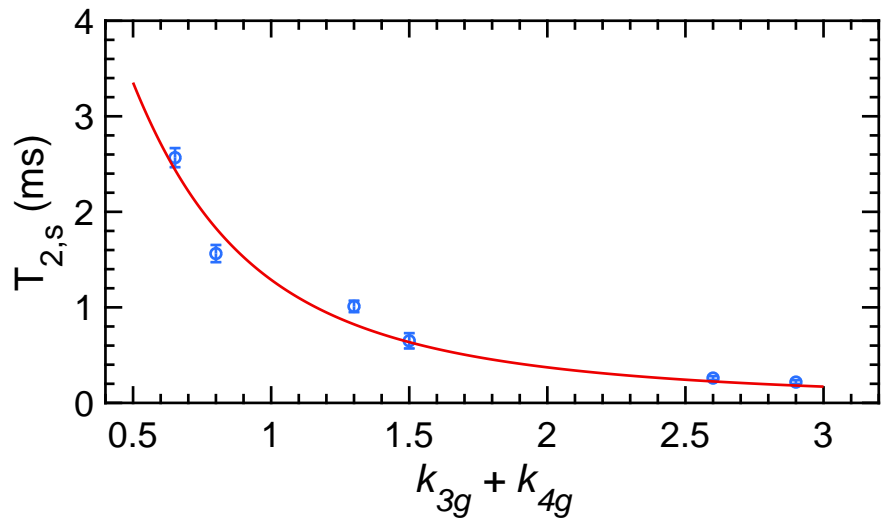}
	\caption{Spin coherence times $T_{2,s}$ of $\ket{4g}\leftrightarrow\ket{3g}$ transition for various  initial populations in $\ket{3g}$ and $\ket{4g}$ spin states (circles) and fit to a model of coherence limited by population lifetimes  (red line, see text).
		}
	\label{fig:T2s_vs_population}
\end{figure}

In a first set of measurements, the optical pumping was performed with the laser set between $\ket{4g}\leftrightarrow\ket{1e}$ and $\ket{3g}\leftrightarrow\ket{1e}$ optical transitions to polarize the spin ensemble into $\ket{1g}$ and $ \ket{2g}$ spin states. The optical pumping, in this set of measurements, is done by scanning the laser over the inhomogeneous broadening during 500~ms to speed up the pumping process and polarize larger spin population. The second laser was used to detect the spin echo signal through RHS detection and was set to $\ket{4g}\leftrightarrow\ket{1e}$ transition. The populations in each state were estimated using separate optical absorption measurements utilizing the previously measured optical branching ratio table (see section \ref{Ratios}). As a result, we observe a substantial increase of the spin coherence time up to 2.5~ms for the strongest polarization of the spin ensemble (Fig.~\ref{fig:T2s_vs_population}).  We note that similar values were measured previously in this sample \cite{Ortu:2018ig}.

On a second stage, the optical pumping was performed with a laser set between the $\ket{1g}\leftrightarrow\ket{4e}$ and the $\ket{2g}\leftrightarrow\ket{3e}$ optical transitions to inverse the polarization and have  higher population of $\ket{3g}$ and $\ket{4g}$ spin states. Additionally the transition $\ket{3g}\leftrightarrow\ket{2e}$ was weakly driven to initialize the spin ensemble to create initial spin polarization for the RHS generation. As a result, a reduction of the spin coherence time up to 0.2~ms was measured (Fig.~\ref{fig:T2s_vs_population}). 

We attribute this behavior to the modification of the spin flip-flop process on $\ket{3g}\leftrightarrow\ket{4g}$ transition, directly limiting the spin coherence time through $\ket{3g}$  and $\ket{4g}$ population lifetimes. Indeed, in contrast to the optical coherence study, the increase of the spin coherence was measured when pumping into both $\ket{3g}$ and $\ket{4g}$. In this case, the optical pumping reduces the flip-flop rate on $\ket{3g}\leftrightarrow\ket{4g}$ transition, by proportionally increasing the cross relaxation between $\ket{1g}\leftrightarrow\ket{2g}$ ground states.  Assuming that the flip-flop probabilities on $\ket{1g}\leftrightarrow\ket{2g}$ and $\ket{3g}\leftrightarrow\ket{4g}$ transition are the same, such population distribution doesn't lead to an overall reduction of the magnetic field noise created by the flip-flops, which could potentially explain the coherence increase. In this situation, the increase of the spin coherence on $\ket{3g}\leftrightarrow\ket{4g}$ transition can be explained by the increase of $\ket{3g}$  and $\ket{4g}$ population lifetimes induced by the reduction of flip-flops  between $\ket{3g}$ and $\ket{4g}$ spin states.

To estimate the spin flip-flop rate we apply the simple coherence time model from the main text (Eq.~(3)). For this we assume that the pure dephasing term is constant for different pumping conditions and the lifetime of the  ground states is limited by the flip-flop process on $\ket{4g}\leftrightarrow\ket{3g}$ transition. 
The last assumption results in quadratic dependence on the population in these two spin states:
\begin{equation}
T_{2,s}^{-1} = R_{ff} k'_{3g}k_{4g}' + \pi \Gamma_{\phi},
\label{eq:T2s_model}
\end{equation}
where $k'_{3g}$ and $k_{4g}' $ are the populations after applying the $\pi/2$ microwave pulse for the spin echo measurement. Since the first $\pi/2$ microwave pulse will average initial populations between $\ket{3g}$ and $\ket{4g}$ spin states we can write $k'_{3g} k'_{4g} = ((k_{3g}+k_{4g})/2)^2$. Initial populations $k_{3g}$, $k_{4g}$ for various optical pumping conditions were measured by fitting the absorption profile taken before applying the microwave sequence.
By fitting the measured coherence times $T_{2,s}$ (Fig. \ref{fig:T2s_vs_population}) using Eq.~(\ref{eq:T2s_model}) we estimate the flip-flop rate to be 0.39~ms for equal population of all spin states, with spin coherence time limit of $(\pi \Gamma_{\phi})^{-1} = 7.2$~ms.

The estimated equilibrium flip-flop process will limit the optical coherence to 0.8~ms which is more than two times bigger than the optical coherence time of 0.3~ms measured without optical pumping. This can be explained by a stronger sensitivity of the optical transition frequency to magnetic field fluctuations coming from the crystalline spin bath, which are modified by the optical pumping.

\section{Discussion}

\subsection{Polarization into a single hyperfine level}

\begin{figure}
\includegraphics[scale=0.7]{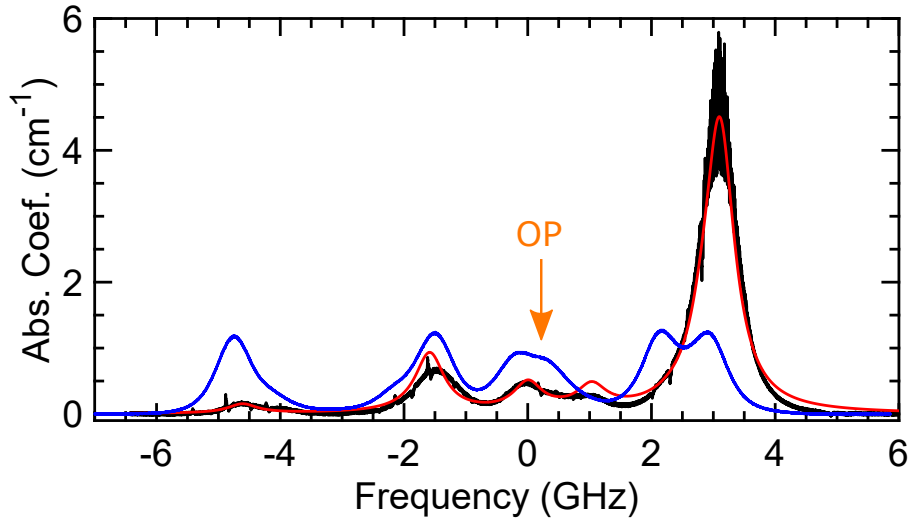}
\caption{Absorption spectrum after 20 s OP for the laser set at +0.22 GHz (black line) and fit (red line). Corresponding normalized populations are: $k_1 = 3.84 \pm 0.02$, $k_2 = 0.04 \pm 0.02$, $k_3 = 0.12 \pm 0.12$, $k_4 = 0.12 \pm 0.04$. Blue line: absorption spectrum at thermal equilibrium. }
\label{pumping_3levels}
\end{figure}

Here the laser is set at +0.22 GHz on the absorption spectrum shown on Fig. \ref{pumping_3levels}. At this frequency, the laser is resonant with some ions in $\ket{2_g}$, $\ket{3_g}$ and $\ket{4_g}$, through the optical transitions $\ket{2_g} \leftrightarrow \ket{2_e}$, $\ket{3_g} \leftrightarrow \ket{3_e}$, $\ket{3_g} \leftrightarrow \ket{4_e}$, $\ket{4_g} \leftrightarrow \ket{3_e}$ and  $\ket{4_g} \leftrightarrow \ket{4_e}$ (see Fig.  \ref{deconvo_eq}b). After 20s of OP duration, the absorption spectrum shown in Fig. \ref{pumping_3levels} is obtained. Corresponding normalized populations are $k_1 = 3.84 \pm 0.02$, $k_2 = 0.04 \pm 0.02$, $k_3 = 0.12 \pm 0.12$, $k_4 = 0.12 \pm 0.04$, which means that 96 $\pm\ 1$ \% of the total population in the volume addressed by the laser has been stored into the $\ket{1_g}$ state. As expected, we can see that the $\ket{1_g} \leftrightarrow \ket{4_e}$ absorption has reached a value almost four times larger than at thermal equilibrium.

\bibliography{Papers}

\end{document}